\documentclass[%
 amsmath,amssymb,
 aps,
]{revtex4-2}

\usepackage{graphicx}
\usepackage{dcolumn}
\usepackage{bm}

\usepackage[dvipsnames]{xcolor}
\usepackage[caption=false]{subfig}
\usepackage{hyperref}
\usepackage{cleveref}
\usepackage{tikz}
\usetikzlibrary{shapes,plotmarks}
\usepackage{rotating}
\usepackage{adjustbox}
\usepackage{etoolbox}
\usepackage{MnSymbol}
\usepackage{float}

\DeclareRobustCommand\full  {\tikz[baseline=-0.6ex]\draw[thick] (0,0)--(0.5,0);}
\DeclareRobustCommand\dotted{\tikz[baseline=-0.6ex]\draw[thick,dotted] (0,0)--(0.5,0);}
\DeclareRobustCommand\dashed{\tikz[baseline=-0.6ex]\draw[thick,dashed] (0,0)--(0.5,0);}

\DeclareMathOperator{\strip}{\mathcal{I}_{\sigma_-,\sigma_+}}
\DeclareMathOperator{\Mellin}{\mathcal{W}}

\DeclareMathOperator*{\app}{\approx}

\definecolor{darkgreen}{RGB}{0.0, 100, 0.0}
\definecolor{purple}{RGB}{128, 0.0, 128}
\definecolor{orange}{RGB}{255, 149, 0.0}
\definecolor{grey}{RGB}{127, 127, 127}

\definecolor{blueplot}{RGB}{12.0, 93, 165}
\definecolor{greenplot}{RGB}{0.0, 185, 69}
\definecolor{orangeplot}{RGB}{255, 149, 0.0}
\definecolor{brownplot}{RGB}{140, 86, 75}
\definecolor{pinkplot}{RGB}{227, 119, 194}

\begin{document}

\preprint{APS/123-QED}

\title{Stability of Stationary Solutions in Acoustic Wave Turbulence}

\author{Guillaume Costa}
\email{guillaume.costa@oca.eu}
\affiliation{Université Côte d’Azur, Observatoire de la Côte d’Azur, CNRS, Laboratoire Lagrange, France}

\author{Giorgio Krstulovic}
\affiliation{Université Côte d’Azur, Observatoire de la Côte d’Azur, CNRS, Laboratoire Lagrange, France}

\author{Sergey Nazarenko}
\affiliation{Université Côte d’Azur, CNRS, Institut de Physique de Nice (INPHYNI), France}

\date{\today}

\begin{abstract}
We study the stability of steady-state solutions of the Wave-Kinetic Equations for acoustic waves. Combining theoretical analysis and numerical simulations, we characterise the time evolution of small isotropic perturbations for both 2D and 3D equilibrium Rayleigh-Jeans and non-equilibrium Kolmogorov-Zakharov solutions. In particular, we show that the stability of these solutions is ensured by different mechanisms.
\end{abstract}

\maketitle

\section{Introduction}
    From ocean surfaces to atmospheric layers, waves are ubiquitous across a wide range of spatial and temporal scales. Understanding their behaviour and interactions is of fundamental interest in many fields of physics for instance, in oceanography, plasma physics~\cite{vedenov1967theory, zaslavskii1967limits}, and nonlinear optics~\cite{dyachenko1992optical, nazarenko2006wave, nazarenko2007freely}. One of the main theoretical frameworks used to study these phenomena is the Weak Wave Turbulence (WWT) theory~\cite{nazarenko2011wave, zakharov2012kolmogorov}, which models the statistical behavior of systems composed of a large number of weakly interacting waves. At the heart of this theory lies the Wave Kinetic Equation (WKE), which governs the evolution of the wave-action spectrum and has been derived for various physical systems, including surface gravity and capillary waves~\cite{krasitskii1990canonical,zakharov1999statistical}, internal waves~\cite{caillol2000kinetic, galtier2003weak}, magnetohydrodynamic (MHD) waves~\cite{goldreich1995toward}, acoustic waves~\cite{zakharov1965weak} etc...
    When driven out-of-equilibrium through forcing and dissipation, as in most real-world scenarios, systems exhibit constant fluxes of conserved quantities. In particular, it has been observed both experimentally~\cite{brazhnikov2001capillary, abdurakhimov2007turbulence, falcon2009capillary} and numerically~\cite{falkovich1988energy, pushkarev2000turbulence, griffin2022energy, zhu2023direct, zhu2023self, labarre2024wave} that, under such conditions, the wave-action spectrum can reach the Kolmogorov-Zakharov (KZ) steady-state, in agreement with predictions of WWT~\cite{zakharov1965weak}. 
    \medbreak\noindent However, the stability of these stationary solutions remains a challenging and largely unresolved question, which has thus far received limited attention. While the observation of these steady-states seems to indicate stability, the propagation of a small perturbation and the conditions under which instability may arise are not yet fully understood. Notably, there are also other WWT systems in which the KZ spectra are not observed, and potential instability is a key suspect in explaining these deviations.
    \medbreak\noindent Only a few studies addressing stability have been carried out to date. Notably, Balk \& Zakharov~\cite{balk1998stability} developed a comprehensive theory for the stability of solutions with respect to small perturbations. Their approach is based on the Mellin transform, formally a Fourier transform in logarithmic coordinates, which leads to a Carleman-type equation. Using methods from complex analysis, they demonstrated that all information regarding stability is encapsulated in a single function: the Mellin function $\Mellin$, which will be discussed in detail later. 

   \medbreak\noindent Besides the work of Balk \& Zakharov, the stability of several systems has been investigated. for instance, Escobedo and collaborators investigated the stability of Bogoliubov waves~\cite{escobedo2011analytical} (related to Bose-Einstein condensates) as well as solutions to the Nonlinear Schrödinger wave kinetic equation (NLS-WKE)~\cite{escobedo2007fundamental}. Following this work, several mathematical studies on the stability of NLS-WKE solutions were published including stability of Rayleigh-Jeans~\cite{escobedo2024instability} and Kolmogorov-Zakharov~\cite{collot2024stability} solutions of the 3D NLS-WKE. It is also worth noting that contemporaneously to Balk and Zakharov’s work, Falkovich~\cite{fal1987stability} analytically investigated the stability of the two-dimensional acoustic KZ spectrum.
    
    \medbreak\noindent In this article, we study the stability of the different steady-state solutions with respect to isotropic perturbations in the context of 2D and 3D acoustic waves. Section~\ref{section:WKE} presents the theoretical background of the WKE and its steady-state solutions. Section~\ref{Section:Numerical} presents the numerical framework used to perform our simulations. Sections~\ref{Section:RJ} \&~\ref{Section:KZ} investigate the stability of the various steady-state solutions of the WKE through both theoretical and numerical analysis.

\section{Theoretical background around the WKE}
    \label{section:WKE}
    For 3-wave processes, the statistical behaviour of waves is governed by the following WKE~\cite{zakharov1965weak, nazarenko2011wave}:
    \begin{equation}
        \begin{split}
             & \partial_t n_{\bm{k}} = 2\pi\int (\mathcal{R}^{\bm{k}}_{\bm{1}, \bm{2}} - 2\mathcal{R}^{\bm{1}}_{\bm{k}, \bm{2}})d\bm{k_1}d\bm{k_2} = S_t[n_{\bm{k}}],  \\
            & \mathcal{R}^{\bm{k}}_{\bm{1}, \bm{2}} = |V^{\bm{k}}_{\bm{1},\bm{2}}|^2\delta^{\bm{k}}_{\bm{1}, \bm{2}}\delta(\omega^{\bm{k}}_{\bm{1},\bm{2}})[n_{\bm{1}}n_{\bm{2}} - n_{\bm{k}}(n_{\bm{1}}+n_{\bm{2}})], \\
            & \delta^{\bm{k}}_{\bm{1}, \bm{2}} = \delta(\bm{k} - \bm{k_1} - \bm{k_2}), \\
            & \delta(\omega^{\bm{k}}_{\bm{1},\bm{2}}) = \delta(\omega_{\bm{k}} - \omega_{\bm{k_1}}-\omega_{\bm{k_2}}),
        \end{split}
        \label{eq:kinetic}
    \end{equation}
    where $k = |\bm{k}|$ is the magnitude of the wave vector, $V^{\bm{k}}_{\bm{1},\bm{2}} \equiv V(\bm{k},\bm{k_1},\bm{k_2})$ is the interaction amplitude and $n_{\bm{i}} \equiv n(\bm{k_i})$ is a shorthand notation for the wave-action. The resonant manifold $\bm{\Sigma}$ is defined as the set of wavevector $(\bm{k}, \bm{k}_1, \bm{k}_2) \in \mathbb{R}_+^{3D}$ for which both wavevector and frequency are conserved:
    \begin{equation*}
        \bm{k} = \bm{k}_1 + \bm{k}_2, \qquad \omega_{\bm{k}} = \omega_{\bm{k}_1} + \omega_{\bm{k}_2},
    \end{equation*}
    where the dispersion relation is usually given by $\omega(\bm{k}) \propto k^{\alpha}$. Equivalently, $\bm{\Sigma}$ is the support of the diracs $\delta^{\bm{k}}_{\bm{1}, \bm{2}}\, \delta(\omega^{\bm{k}}_{\bm{1}, \bm{2}})$, and integrals involving this delta product are referred to as integrals over the resonant manifold~$\bm{\Sigma}$.

    \medbreak\noindent Eq.~\ref{eq:kinetic} conserves the total energy $E$, defined as
    \begin{equation*}
        E = \int \omega_{\bm{k}} n_{\bm{k}}d\bm{k},
    \end{equation*}
    implying the possibility of an energy flux in the system.
    
    \subsection{WKE for acoustic waves}
        \medbreak\noindent In this article, we study the stability of isotropic steady-state solutions in the context of acoustic waves, whose behaviours are described by the WKE~\eqref{eq:kinetic} with interaction coefficient $V^{\bm{k}}_{\bm{1}, \bm{2}} = V_0\sqrt{k k_1 k_2}$. The wave frequency is typically given by the linear dispersion relation $\omega_k = c_sk$. However, this linear relation leads to singular behavior in the WKE. Indeed, let $n_k$ be an isotropic solution of Eq.~\eqref{eq:kinetic} and rewrite the $D + 1$ resonance conditions as:
        \begin{equation*}
            \delta^{\bm{k}}_{\bm{1}, \bm{2}}\delta(\omega^{\bm{k}}_{\bm{1},\bm{2}}) = \delta(\bm{\mathcal{F}}(\bm{\mathcal{E}}_D, \overline{\bm{\mathcal{E}}_D}))
        \end{equation*}
        where $\bm{\mathcal{F}}(\bm{\mathcal{E}}_D, \overline{\bm{\mathcal{E}}_D})$ encodes the resonance conditions. In particular, $\bm{\mathcal{E}}_D$ are $D+1$ variables removed by integrating over the Dirac-deltas and $\overline{\bm{\mathcal{E}}_D}$ are the remaining $2D-1$ variables. A natural choice for $\bm{\mathcal{E}}_D$ is given by:
        \begin{equation*}
            \bm{\mathcal{E}}_D = \begin{cases}
                   (k_2, \theta_1,\theta_2),  & D = 2, \\
                   (k_2, \theta_1,\theta_2, \phi_1), & D = 3.
            \end{cases}
        \end{equation*}
        Using the standard identity for delta functions under a change of variables, one obtains:
        \begin{equation*}
                       \delta(\bm{\mathcal{F}}(\bm{\mathcal{E}}_D, \overline{\bm{\mathcal{E}}_D})) = \sum_{\bm{\mathcal{E}}_D^*} \frac{\delta(\bm{\mathcal{E}}_D - \bm{\mathcal{E}}_D^*)}{|\det\bm{\nabla}_{\bm{\mathcal{E}}_D} \bm{\mathcal{F}}(\bm{\mathcal{E}}_D^*, \overline{\bm{\mathcal{E}}_D^*})|},
        \end{equation*} 
        where the roots of $\bm{\mathcal{F}}(., \overline{\bm{\mathcal{E}}_D})$ and $\bm{\nabla}_{\bm{\mathcal{E}}_D}\bm{\mathcal{F}}$ correspond to the Jacobian of $\bm{\mathcal{F}}$ with respect to the ${\bm{\mathcal{E}}_D}$ variables.

        \medbreak\noindent Substituting into the WKE~\ref{eq:kinetic} yields:
        \begin{equation*}
            \begin{split}
                \partial_t n_k \propto  \Bigg(
                &\sum_{\bm{\mathcal{E}}_D^*}\int_0^k |V^k_{1,2}|^2 \frac{\delta(\bm{\mathcal{E}}_D - \bm{\mathcal{E}}_D^*)}{|\det\bm{\nabla}_{\bm{\mathcal{E}}_D} \bm{\mathcal{F}}(\bm{\mathcal{E}}_D^*, \overline{\bm{\mathcal{E}}_D^*})|}
                \big(n_{1} n_{2} - n_k(n_{1} + n_{2})\big) (k_1 k_2)^{D-1} \, d\bm{k_1} d\bm{k_2} \\
                - 2 &\sum_{\bm{\mathcal{E}}_D^*}\int_k^\infty |V^1_{k,2}|^2 \frac{\delta(\bm{\mathcal{E}}_D - \bm{\mathcal{E}}_D^*)}{|\det\bm{\nabla}_{\bm{\mathcal{E}}_D} \bm{\mathcal{F}}(\bm{\mathcal{E}}_D^*, \overline{\bm{\mathcal{E}}_D^*})|}
                \big(n_k n_{2} - n_{1}(n_k + n_{2})\big) (k_1 k_2)^{D-1} \, d\bm{k_1} d\bm{k_2}
                \Bigg).
            \end{split}
        \end{equation*}
        For a linear dispersion relation, the resonant manifold becomes degenerate at all point as $|\det \bm{\nabla} \bm{\mathcal{F}}| = 0$. This results in a divergence of the collision integral, making the equation ill-posed. In 3D, this singularity can be removed through angle integration under isotropic assumptions. However, the 2D resonant manifold remains degenerate and the collision integral diverges. It is then necessary to regularize the integral, which can be done~\cite{zakharov2012kolmogorov} by either:
        \begin{enumerate}
            \item perturbing the exponent of the dispersion relation $\omega(k) = c_sk^{\alpha + \epsilon}$, $\epsilon \ll 1$,
            \item introducing a small nonlinear dispersion such that the modified dispersion relation reads $\omega(k) = c_sk(1+a^2k^2)$, where $a$ is the dispersive length.
        \end{enumerate}
        While both approaches lead to an effective regularization of the collision integral~\cite{zakharov2012kolmogorov}, the addition of a small nonlinearity has more physical meaning. For instance, the Gross-Pitaevskii equation, a well established model for Bose-Einstein condensates, behaves in the large-scale limit as weakly dispersive sound waves with $a = \xi / 2$, $\xi$ being the healing length~\cite{griffin2022energy}. Hence, we consider the modified dispersion relation (ii), taken in the weakly dispersive limit $ak \ll 1$.
        
        \medbreak\noindent Applying the regularization and integrating over both the resonant conditions and remaining angles, under the isotropic assumption, lead to the following equation:
        \begin{equation*}
            \begin{aligned}
                \partial_t n_k \propto
                \Bigg(&\int_0^k \frac{|V^k_{1,2}|^2}{c_s[1 + (3 a k_2)^2] \Delta_{k12}} \delta(k-k_1-k_2)\big(n_1 n_2 - n_k(n_1 + n_2)\big) (k_1 k_2)^{D-1} \, dk_1dk_2 \\
                - 2 &\int_k^\infty \frac{|V^1_{k,2}|^2}{c_s[1 + (3 a k_2)^2] \Delta_{1k2}} \delta(k_1-k-k_2) \big(n_k n_2 - n_1(n_k + n_2)\big) (k_1 k_2)^{D-1}\, dk_1dk_2 \Bigg),
            \end{aligned}
        \end{equation*}
        where $k_2 \neq |k-k_1|$ is now a function of $(k, k_1)$ obtained using Cardano formulas, and $\Delta_{k12}$ is given by
        \begin{align*}
            \Delta_{k12} = \begin{cases}
                \frac{1}{2} \sqrt{
                    2(k^2 k_1^2 + k^2 k_2^2 + k_1^2 k_2^2) 
                    - k^4 - k_1^4 - k_2^4}, & D = 2, \\
                k k_1 k_2, & D = 3.
            \end{cases}
        \end{align*}
    
        \medbreak\noindent Upon taking the small-dispersion limit $a k \ll 1$, one obtains:
        \begin{align*}
            &\Delta_{k12} = \begin{cases}
                \sqrt{6} a k k_1 k_2 + \mathcal{O}(a^3), & D = 2, \\
                k k_1 k_2, & D = 3,
            \end{cases} \\
            &k_2 = |k - k_1| + \mathcal{O}(a^2),
        \end{align*}
        effectively leading to an integrable action, agreeing with the analysis of~\cite{griffin2022energy}.
        
        \medbreak\noindent The simplified acoustic WKE in dimension $D$ finally reads:
        \begin{equation}
            \begin{aligned}
                \partial_t n_k = C_D
                \Bigg(&\int_0^k [k_1(k - k_1)]^{D-1} [n_1 n_2 - n_k(n_1 + n_2)] \, dk_1 \\
                - 2 &\int_k^\infty [k_1(k_1 - k)]^{D-1} [n_k n_2 - n_1(n_k + n_2)] \, dk_1 \Bigg),
            \end{aligned}
            \label{eq:SimpWKE}
        \end{equation}
        where $C_D$ is a constant depending on the dimension given by:
        \begin{equation}
            C_D = 
        \begin{cases}
            \dfrac{4 \pi V_0^2}{c_s \sqrt{6} a}, & D = 2, \\
            \dfrac{4 \pi^2 V_0^2}{c_s}, & D = 3.
        \end{cases}
        \label{eq:Cd}
        \end{equation}
        These expressions were originally obtained in~\cite{griffin2022energy} for the 2D case~\footnote{Note that the expression derived in~\cite{griffin2022energy} appears to lack a factor of 2 in the prefactor.} and in~\cite{zakharov1970spectrum} for the 3D case. Note that in the 3D case, the equation no longer depends on the dispersive length $a$. However, the prefactor differs by a factor of 2 between the standard and regularized integrals~\cite{zhu2024turbulence}, a discrepancy attributed to the non-commutativity of the limit and the integral.
    
    \subsection{Steady-state solutions of the WKE}
        \label{section:Steady_state}
        Steady-state solutions of the WKE~\ref{eq:SimpWKE} are characterized by a vanishing collision integral $S_t[n_k] \equiv 0$. To show the existence of such solutions, we first rewrite the collision integral in a more convenient form:
        \begin{equation}
            \begin{aligned}
                \partial_tn_{k} &= S_t[n_k] \equiv C_D\int_{0}^{k} \big(k_1k_2\big)^{D-1}\big[\mathcal{N}^k_{1,2} -\mathcal{N}^1_{k,2} -\mathcal{N}^2_{1,k}\big]dk_1dk_2, \\ 
               \mathcal{N}^k_{1,2} &=
               \delta(k-k_1-k_2)[n_1n_2 - n_k(n_1+n_2)] 
            \end{aligned}     
            \label{eq:SteadyWKE}
        \end{equation}
        We now seek isotropic, powerlaw steady-state solutions of the form $n^0_k = B k^{-\mu}$, where $(B,\mu)$ are two constants. A possible steady-state solution is immediately recovered solving $\delta(\omega^{k}_{1,2}) = 0$, corresponding to a fluxless, thermal equilibrium solution with a Rayleigh-Jeans (RJ) spectrum $n^0_k \propto \omega_k^{-1} = \dfrac{1}{c_sk}$. 

        \medbreak\noindent Other steady-state solutions can be found by applying the Kraichnan–Zakharov~\cite{zakharov1966weak, zakharov1972collapse} transformation to the second and third terms of the WKE~\ref{eq:SteadyWKE}, leading to:

        \begin{equation*}
            \begin{aligned}
                S_t[n_k] &= C_D \, B^2 \, I(\mu) \, k^{2D - 1 - 2\mu}, \\
                I(\mu) &= \int_0^1 \big[q(1 - q)\big]^{D - 1} 
                \left[1 - q^{-y} - (1 - q)^{-y}\right]
                \left[q^{-\mu}(1 - q)^{-\mu} - q^{-\mu} - (1 - q)^{-\mu} \right] \, dq, \\
                y &= 3D - 1 - 2\mu.
            \end{aligned}
            \label{eq:AdimColl}
        \end{equation*}
        Steady-state solutions are then obtained by identifying the value of $\mu$ for which the integrand vanishes, i.e. $y = -1$, leading to the KZ spectrum with constants $(B,\mu)$ given by~\cite{zhu2024turbulence, griffin2022energy}:
        \begin{equation*}
            \begin{cases}
                D = 2: \quad B = \dfrac{4}{3\pi^2}\sqrt{\dfrac{2aP}{c_s}}, \quad \mu = 3, \\[6pt]
                D = 3: \quad B = \dfrac{ \sqrt{6P}}{3\pi c_s \sqrt{32\pi(\pi + 4\ln 2 - 1)}}, \quad \mu = \dfrac{9}{2},
            \end{cases}
        \end{equation*}
        where $P$ corresponds to the energy flux. Unlike the RJ spectrum, the KZ spectrum corresponds to a non-equilibrium steady-state, with non-zero flux, usually introduced through energy injection and dissipation.

        \medbreak\noindent While we have shown that the WKE~\ref{eq:SimpWKE} admits isotropic powerlaw steady-state solutions, it remains necessary for the collision integral (i.e. for $I(\mu)$) to be convergent for powerlaw solutions $k^{-\mu_*}$, $|\mu_* - \mu| \ll 1$. We refer to the $\mu$-interval of convergence of $I(\mu)$ as the \textit{locality window} and request $I(\mu) \in \mathcal{C}^1(\mathbb{R}, \mathbb{R})$ so as to define the energy flux.
        
        \medbreak\noindent Although the collision integral is trivially convergent for the RJ spectrum as its integrand vanishes identically, any positive (resp. negative) deviation in the exponent is associated with a divergence of the lower (resp. upper) bound of the integral. Consequently, the study of RJ solutions can only be achieved with the addition of a cutoff which, in turn, ensures a finite energy density. This restriction does not apply to KZ solutions, as one can show that the integral $I(\mu)$ converges for all $\mu \in \big]2D-2, D+2\big[$ (see Appendix~\ref{section:App_Loca}).

    \subsection{Linearized WKE for small perturbations}
        To analyze the stability of these steady-state solutions, following~\cite{balk1998stability}, we consider a small isotropic perturbation of the form $n_{k} = n_k^0(1 + A_{k})$, $A_{k} \ll 1$. Substituting $n_{k}$ into Eq.~\ref{eq:kinetic} and discarding higher-order terms yields the linearized evolution equation for the perturbation:
        \begin{equation}
            \begin{split}
                \frac{\partial  A_{k}}{\partial t} &= \frac{1}{n_k^0}\int \big(k_1k_2\big)^{D-1}\big(\mathcal{A}_{1,2}^{k}-2\mathcal{A}_{k,2}^{1}\big)\,dk_1\,dk_2, \\
                \mathcal{A}_{1,2}^{k} &= C_D\delta^k_{1, 2}
                \big[A_1n_1^0(n_2^0 - n_k^0) + A_2n_2^0(n_1^0 - n_k^0) - A_kn_k^0(n_1^0 + n_2^0)\big],
            \end{split}
            \label{eq:linearized}
        \end{equation}
        where we recall that $n_k^0 = B k^{-\mu}$ is an isotropic steady-state solution of the WKE.
        
        \medbreak\noindent In the most general case, Eq.~\ref{eq:linearized} can be expressed using a linear operator $\mathcal{L}$, defined through a kernel $ \mathcal{U} $, as follows:
        \begin{equation*}
            \begin{aligned}
                \frac{\partial  A_{\bm{k}}}{\partial t} &= \mathcal{L}f(\bm{k}), \\
                \mathcal{L}f(\bm{k}) &= \int\mathcal{U}(\bm{k}/\bm{k'})f(\bm{k'})\,d\bm{k'}.
            \end{aligned}
        \end{equation*}
        The existence of such an operator ensures that any perturbation can be decomposed into the spherical harmonic basis, leading to a set of uncoupled equations. This property is fundamental as it allows for isotropic and anisotropic cases to be treated separately. 
        
        \medbreak\noindent While a general theory of stability has been developed by Balk \& Zakharov~\cite{balk1998stability} for both isotropic and anisotropic cases, it fails to apply to anisotropic perturbations in the acoustic case. Indeed, to obtain an integrable action, one needs to add a small dispersion into the dispersion relation. 
        While anisotropic perturbations can be decomposed at leading order into a superposition of angle-independent perturbations aligned along different directions, the next to leading order is associated with potentially divergent angular gradients. Thus implying the necessity of studying the next to leading order. Yet, the addition of a small disperson breaks the homogeneity of the integrand at any order higher than one, thereby preventing the derivation of a Carleman equation, which is central to the theory of Balk \& Zakharov.
        
        \medbreak\noindent In this work, we therefore focus on leading-order perturbations namely isotropic perturbations .

\section{Numerical details}
    \label{Section:Numerical}
      \begin{figure}[!htb]
            \centering
            \adjustbox{valign=t}{
                 \begin{minipage}{0.49\textwidth} 
                \includegraphics[width=\textwidth]{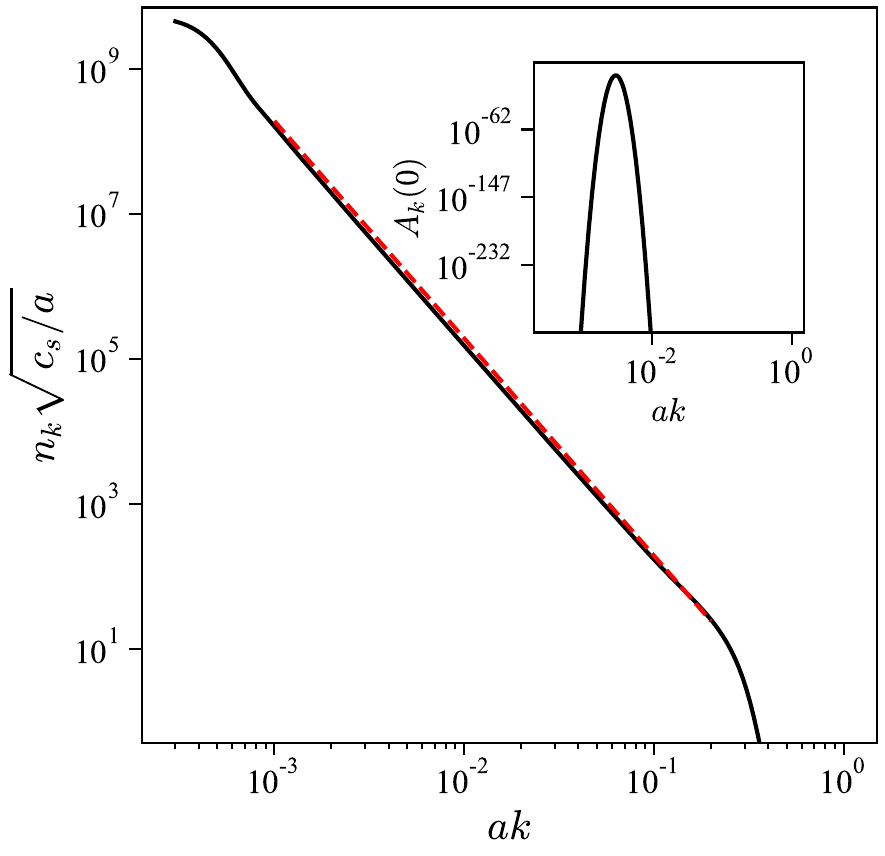}
                \end{minipage}}
            \hfill  
            \adjustbox{valign=t}{\begin{minipage}{0.49\textwidth}
                \includegraphics[width=\textwidth]{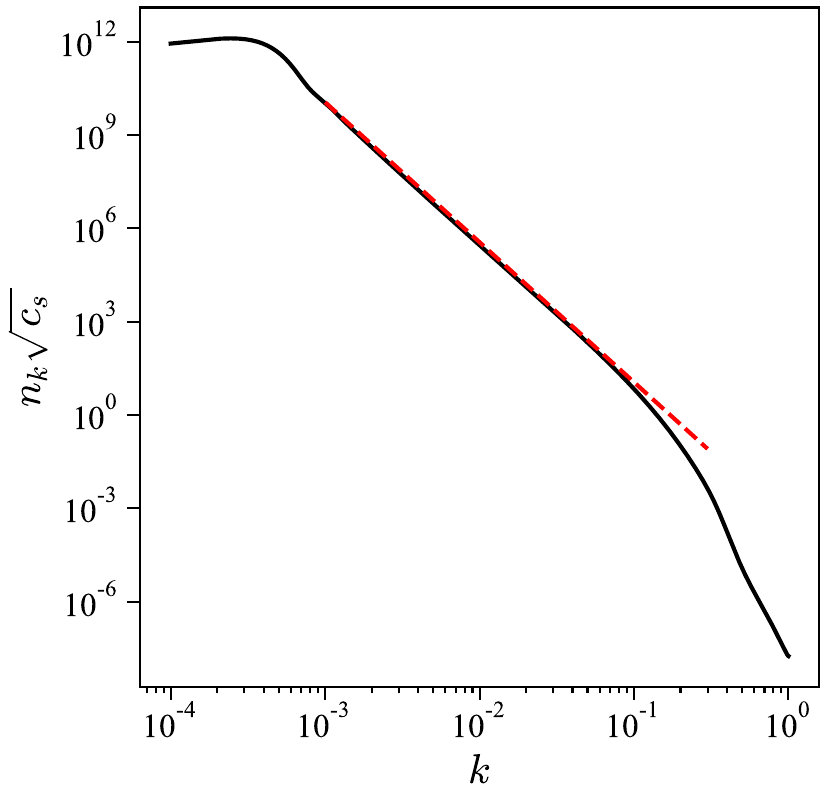}
                \vfill
                \end{minipage}} 
            \put(-505, 10){\resizebox{14pt}{!}{\textbf{(a)}}}
            \put(-248, 10){\resizebox{14pt}{!}{\textbf{(b)}}}
            \put(-395, 10){2D Acoustics}
            \put(-133, 10){3D Acoustics}
            \caption{KZ spectrum for acoustic waves. The solid black lines correspond to numerical simulations while red dashed lines represent the theoretical spectra $n_k = B k^{-\mu}$. \textbf{(a)} 2D case, $B = \dfrac{4}{3\pi^2} \sqrt{\dfrac{2aP}{c_s}}$~\cite{griffin2022energy}, $\mu = 3$. The inset present the initial spectral density of a small isotropic perturbation $A_k(0)$. \textbf{(b)} 3D case, $ B = \dfrac{ \sqrt{6P}}{3\pi c_s \sqrt{32\pi(\pi + 4\ln 2 - 1)}}$~\cite{zhu2024turbulence}, $\mu = 9/2$. Where $P$ corresponds to the energy flux.}
            \label{fig:KZ}
    \end{figure}
    \noindent We perform numerical simulations of the WKE~\ref{eq:SimpWKE} using the WavKinS package~\cite{krstulovic2025wavkins}. The WKE is solved on a logarithmic grid, with wave numbers following a geometric progression $k_n = k_0\lambda^n$. The grid parameter $\lambda$ is fixed by the resolution such that $\lambda(N, k_0, k_{\max}) = (k_{\max} / k_0)^{1/N}$. As previously mentioned, the WKE admits two classes of steady-state solutions depending on the presence of external excitations. To model such effects, we introduce large-scale forcing $f(k)$ and damping $\mathcal{D}(k)$, so that the WKE becomes:
    \begin{equation*}
        \partial_t n_k = S_t[n_k] + f(k) - \mathcal{D}(k),
    \end{equation*}
    where the damping operator is defined as 
    \begin{equation*}
        \mathcal{D}(k) = \nu\left(\dfrac{k}{k_\nu}\right)^\gamma n_k,
    \end{equation*}
    where $k_\nu$ denotes the equivalent of the Kolmogorov scale, $\nu$ is a damping coefficient and $\gamma \in\mathbb{R}$ is an adjustable parameter controlling the dissipation rate

    \medbreak\noindent As stated by the H-theorem, in the absence of external excitations $(f,\nu) \equiv (0,0)$ and in the presence of a cutoff, the system evolves toward a maximum entropy steady-state corresponding to RJ (thermal equilibrium) solutions.  On the other hand, when driven out-of-equilibrium through non-zero excitation $(f,\nu) \not\equiv (0,0)$, the system develops the KZ spectrum. 
    
    \medbreak\noindent For KZ solutions, two types of damping are used: (i) regular viscous dissipation $\gamma = 1$ (ii) hyper-viscous dissipation $\gamma = 4$. The hyper-viscous dissipation allows us to extend, as far as possible ($\sim 3$ decades), the inertial range while keeping reasonable resolutions. This enables the study of the fast propagating 2D perturbation that would otherwise reach the dissipative range too quickly. Figure~\ref{fig:KZ} presents the various KZ spectra.

\section{Equilibrium solutions - Rayleigh-Jeans spectra}
    \label{Section:RJ}
    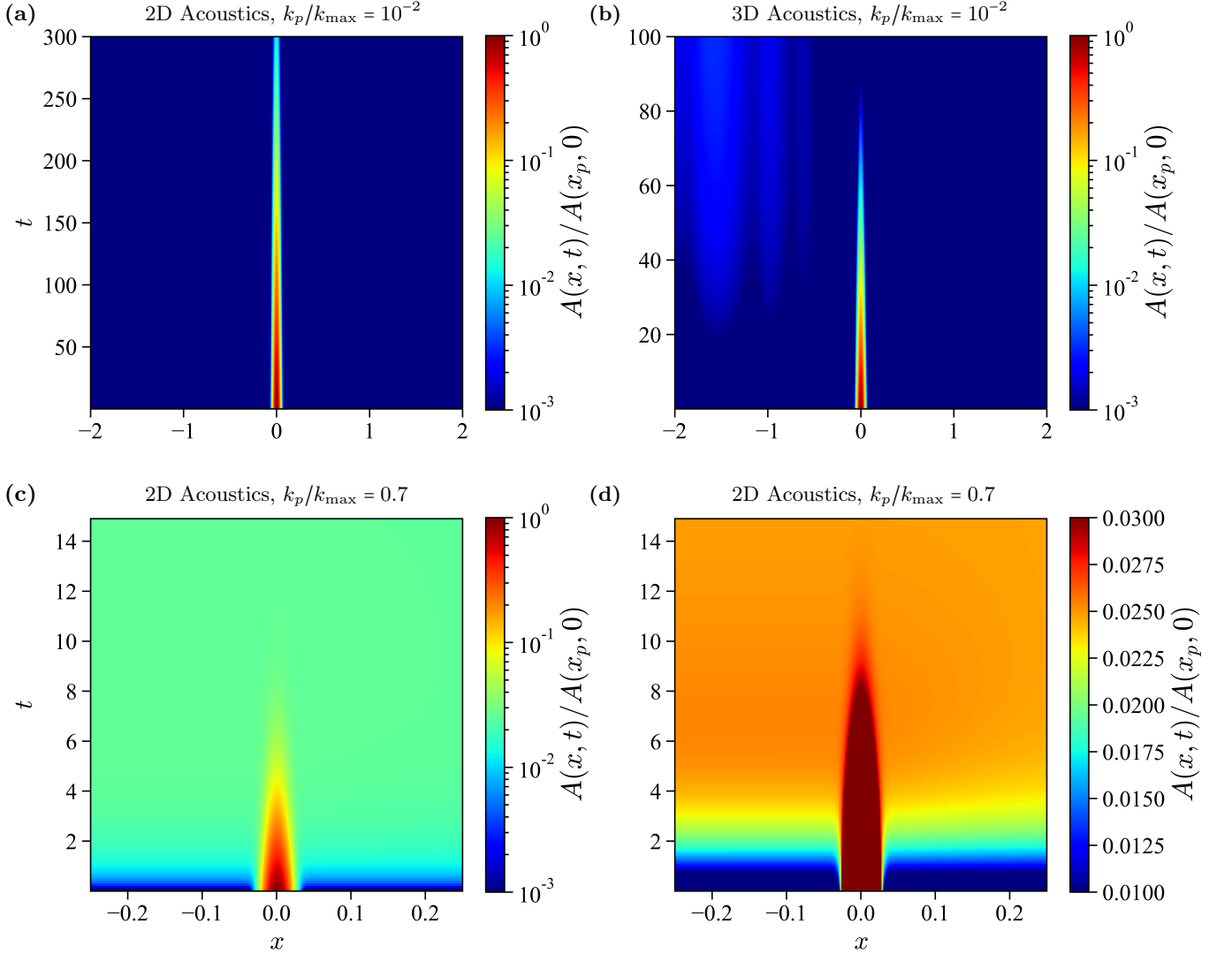
\begin{figure}[!htb]
        \centering
        \includegraphics[width=1.\columnwidth]{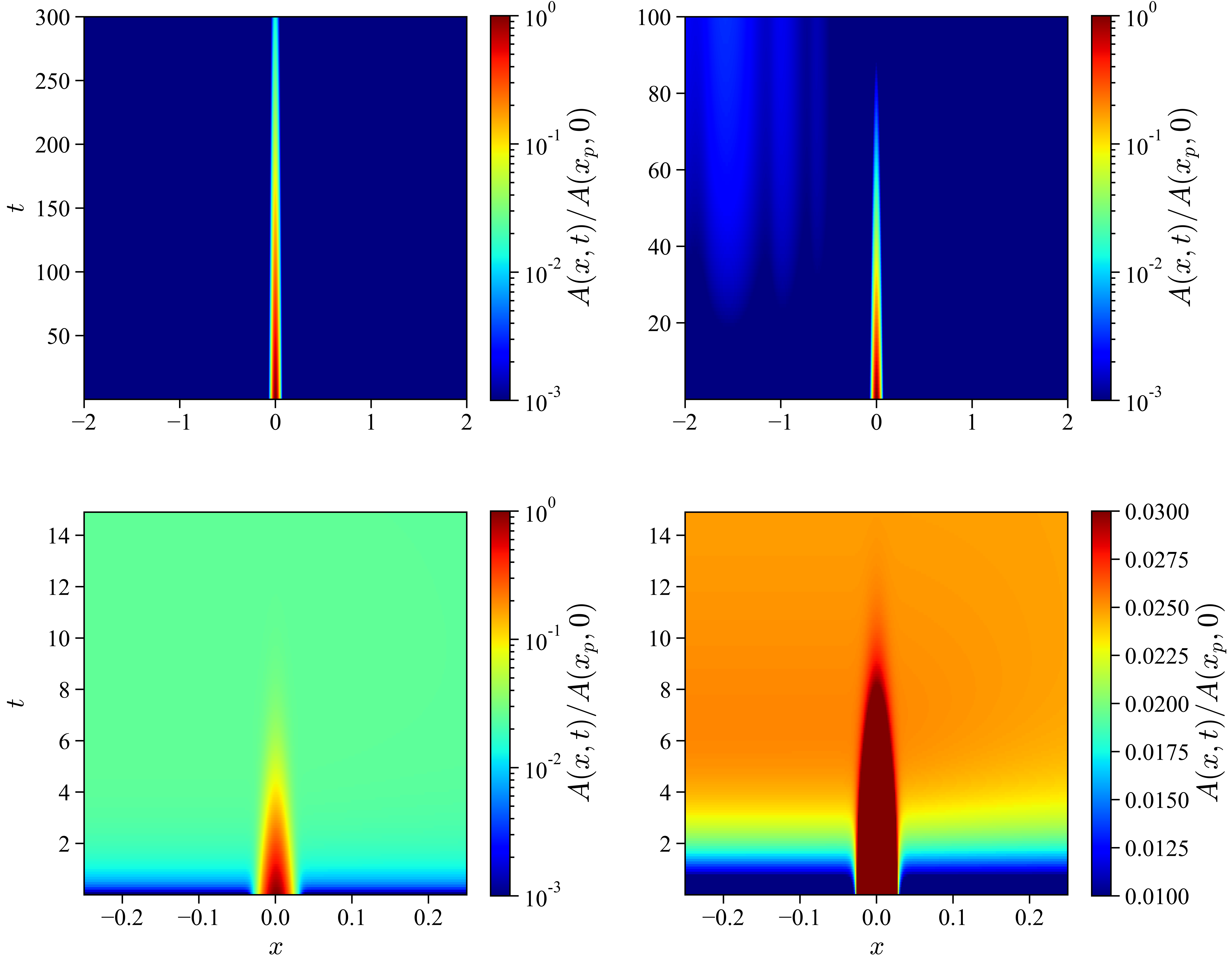}
        \put(-512, 400){\resizebox{14pt}{!}{\textbf{(a)}}}
        \put(-263, 400){\resizebox{14pt}{!}{\textbf{(b)}}}
        \put(-512, 195){\resizebox{14pt}{!}{\textbf{(c)}}}
        \put(-263, 195){\resizebox{14pt}{!}{\textbf{(d)}}}
        \put(-452, 400){2D Acoustics, $k_p/k_{\max} = 10^{-2}$}
        \put(-202, 400){3D Acoustics, $k_p/k_{\max} = 10^{-2}$}
        \put(-452, 195){2D Acoustics, $k_p/k_{\max} = 0.7$}
        \put(-202, 195){2D Acoustics, $k_p/k_{\max} = 0.7$}

        \caption{Heatmaps of the perturbation for RJ solutions. \textbf{(a)} 2D case with $k_p/k_{\max} = 10^{-2}$. \textbf{(b)} 3D case with $k_p/k_{\max} = 10^{-2}$.
        \textbf{(c)} 2D case with $k_p/k_{\max} = 0.7$. \textbf{(d)} Same data as \textbf{(c)} shown with a saturated linear color scale. The \underline{top row} illustrates similar behavior for both 2D and 3D simulations, characterized by a slow decay of the perturbation peak. The \underline{bottom row} highlights that, for larger values of $k_p/k_{\max}$, the decay is significantly faster and associated with a clear broadening of the perturbation.}
        \label{fig:RJ_heatmaps}
    \end{figure}
       
    \noindent The simplest stationary solution of the kinetic equation~\ref{eq:SimpWKE} is obtained by imposing a vanishing flux, i.e., by setting the integrand to be zero at every point on the resonant manifold $\bm{\Sigma}$. This leads to a wave-action spectrum of the form $n_k \propto 1/\omega(k) = \dfrac{1}{c_sk}$, which corresponds to thermal equilibrium. However, this solution lies outside the locality window~\cite{griffin2022energy}, and as a result, its stability cannot be analyzed analytically without introducing an ultraviolet (UV) cutoff $k_{\max}$ (see Section~\ref{section:Steady_state}).
   
    \medbreak\noindent Numerical simulations show that the dimension (2D vs 3D) has little effect on the overall phenomenology (Figure~\ref{fig:RJ_heatmaps}a-b). Instead, the behavior of the perturbation is primarily governed by its localization $k_p/k_{\max}$. For small values of this ratio, the perturbation remains sharply peaked for long time, exhibiting a slow decay (2D - Figure~\ref{fig:RJ_heatmaps}a, 3D Figure~\ref{fig:RJ_heatmaps}b). On the other hand, increasing $k_p/k_{\max}$ leads to a quicker decay and significant broadening of the peak (2D case  Figure~\ref{fig:RJ_heatmaps}c-d). These results suggest the existence of a characteristic time scale $t'$ such that $\forall t < t'$, the perturbation remains localized. 

    \medbreak\noindent We now consider Eq.~\ref{eq:linearized} with the addition of an UV-cutoff $k_{\max}$. Guided by the numerical results (Figure~\ref{fig:RJ_heatmaps}), we introduce the following ansatz for the localized perturbation: $A_k(t) \equiv A(t)\delta(k-k_p)$ and look for an evolution equation for $A(t)$. Applying the methodology of Appendix~\ref{section:Ansatz_RJ}, one obtains, at leading order, the following equation:
    \begin{equation*}
        \frac{dA(t)}{dt} \approx -\alpha(k_p)A(t),
    \end{equation*}
    \noindent which can be explicitely solved leading to an exponential damping of the perturbation $A(t) = A(0)e^{-\alpha t}$, where $\alpha$ is given by:
    \begin{align}
        \label{eq:alpha}
            \alpha(k_p) = \begin{cases}
            \dfrac{2\ V_0^2E^{RJ}}{c_s^2\sqrt{6}a} \dfrac{k_p}{k_{\max}}(2-\dfrac{k_p}{k_{\max}}), & D = 2, \\
            \dfrac{\pi V_0^2E^{RJ}}{2c_s^2}k_p\left[
        \left(\dfrac{k_p}{k_{\max}}\right)^3
        + 2 \left(\dfrac{k_p}{k_{\max}} - 1\right)^2 \left(\dfrac{k_p}{k_{\max}} + 2\right)
        \right],& D = 3.
        \end{cases}
    \end{align}
     
     \begin{figure}[!htb]
          \centering
          \includegraphics[width=1.\columnwidth]{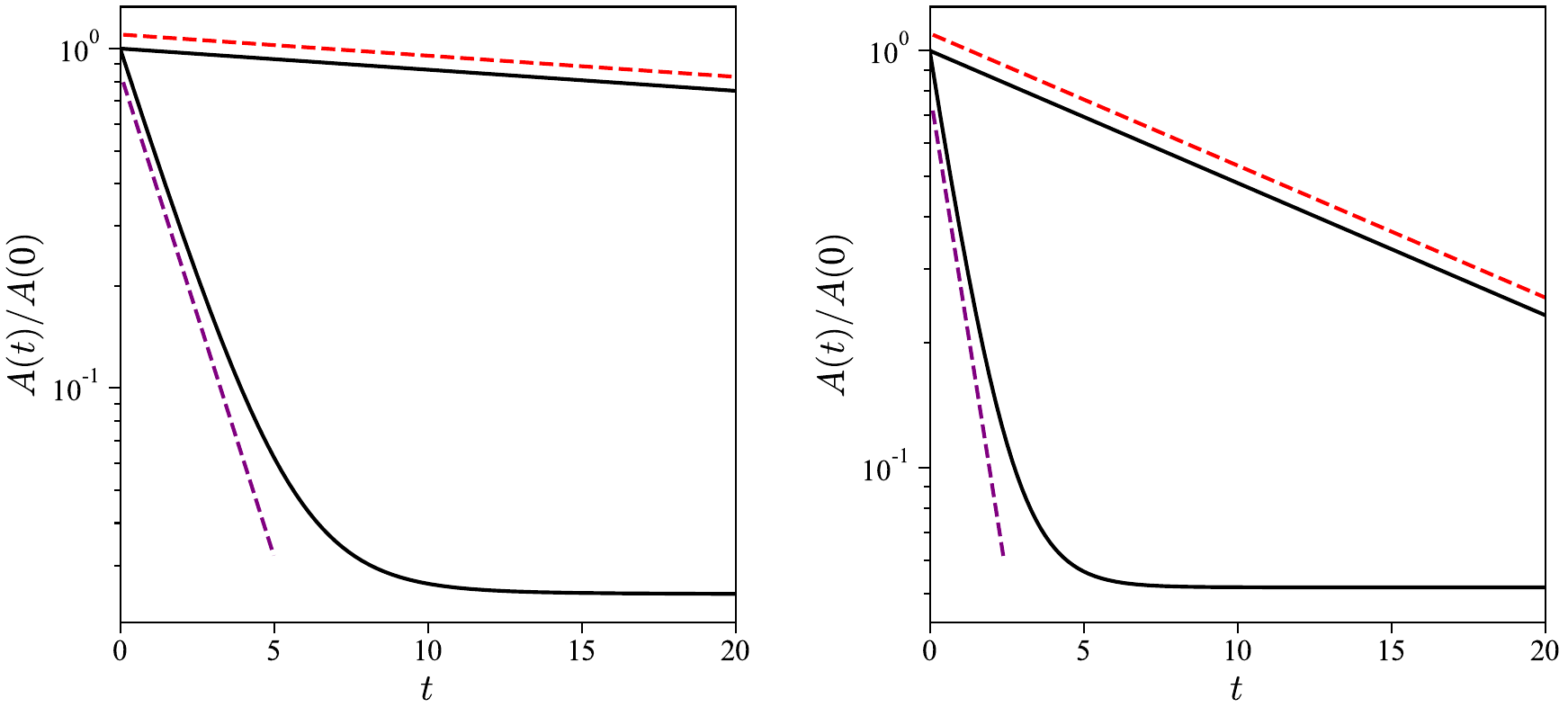}
            \put(-510, 235){\resizebox{14pt}{!}{\textbf{(a)}}}
            \put(-246, 235){\resizebox{14pt}{!}{\textbf{(b)}}}
            \put(-400, 235){2D Acoustics}
            \put(-137, 235){3D Acoustics}
          \caption{Time evolution of the amplitude of a narrow Gaussian perturbation, centered in $k_p$, for the Rayleigh-Jeans solution. Colored dashed lines correspond to theoretical predictions $A(t) = A(0)e^{-\alpha(k_p) t}$, where $\alpha$ is given in Eq.~\ref{eq:alpha}. $\color{red}{\dashed{}}$ $k_p / k_{\max} = 10^{-2}$, $\color{purple}{\dashed{}}$ $k_p /k_{\max} = 0.7$. \textbf{(a)} 2D case. \textbf{(b)} 3D case.}
          \label{fig:RJ}
    \end{figure}
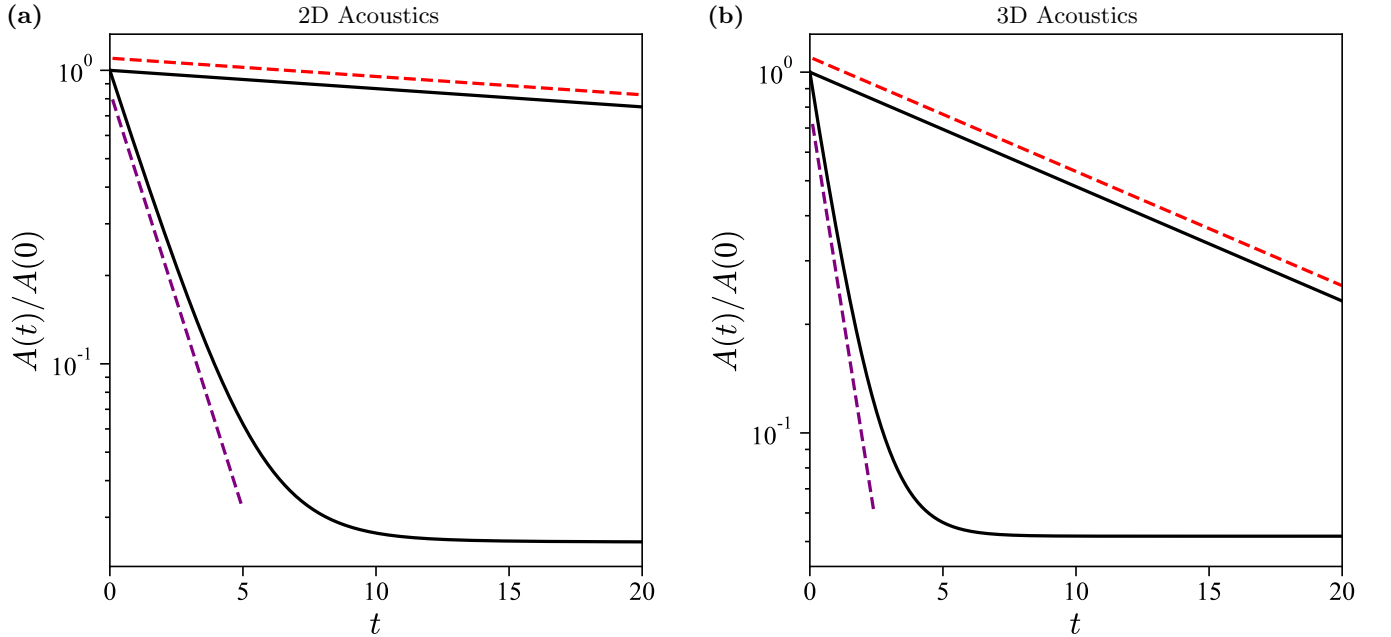
    \noindent Figure~\ref{fig:RJ} presents the time evolution of the amplitude of the perturbation for two different ratio $k_p/k_{\max}$ highlighting good agreement with the theoretical predictions of Eq.~\ref{eq:alpha}. Note that, for long time, the dirac ansatz is no longer valid as the pertubation broadens leading to a new equilibrium state.

    \medbreak\noindent The RJ solutions are thus stable with respect to isotropic perturbations, regardless of the dimension, as $\alpha(k_p) > 0 \; \forall (D, k_p)$.

    \medbreak\noindent While RJ distributions represent the simplest class of steady-state solutions, they only describe fluxless systems in thermodynamic equilibrium. However, most physical systems of interest involve energy and wave-action injection and dissipation, resulting in non-zero fluxes. Therefore, we now turn our attention to the case of non-vanishing fluxes, which are more representative of realistic, non-equilibrium conditions.

\section{Non-equilibrium solutions - Kolmogorov-Zakharov spectra}
    \label{Section:KZ}
    Stationary solutions of Eq.~\ref{eq:kinetic} with non-zero flux correspond to the KZ solutions which exponents $\mu$ depend on the dimension of the system (see Figure~\ref{fig:KZ}). Contrary to the RJ solutions, the acoustic KZ solutions always lie inside the locality window~\cite{griffin2022energy} regardless of the dimension.
    
    \subsection{Carleman equation and Mellin function}
        First, for any complex-valued function $f$, we define its Mellin transform, $\mathcal{M}[f]$, as
        \begin{equation*}
            \mathcal{M}[f](s) = \int_0^{+\infty} k^{s-1} f(k) \, dk, \; \forall s \in \mathbb{C}.
        \end{equation*}
        This transform can then be applied to Eq.~\ref{eq:linearized} to obtain a Carleman type equation~\cite{balk1998stability}:
        \begin{equation}
            \begin{aligned}
                &\partial_t \hat{A}(s + h) = \tau^{-1}\mathcal{W}(s)\hat{A}(s), \\
                &\Mellin(s) = \mathcal{M}[\mathcal{U}] \equiv\int_{-\infty}^{\infty} \mathcal{U}(k)k^{s-1}dk, \\
                &\hat{A}(s) = \mathcal{M}[A] \equiv \int_{-\infty}^{+\infty} A(k,t)k^{s-1}dk,
            \end{aligned}
            \label{eq:Carleman}
        \end{equation}
        where $h = 1 - d/2$ is the homogeneity degree of the collision integral, $\tau$ is a characteristic time scale depending on the dimension $D$ and $\mathcal{W}$ is the Mellin function, containing all the information about the stability of the system. For acoustic systems, the Mellin function can be explicited as:
        
        \begin{equation}
            \begin{split}
                    &\Mellin(s) = \mathcal{}\int_0^1 \big[q(1-q)\big]^{D-1}P_{s,\mu}(q)dq - 2\mathcal{}\int_1^\infty \big[q(q-1)\big]^{D-1}Q_{s,\mu}(q)dq, \\
                    &P_{s,\mu}(q) = \, q^{-s-\mu}((1-q)^{-\mu} - 1) + (1-q)^{-s-\mu}(q^{-\mu} - 1) - q^{-\mu} - (1-q)^{-\mu},  \\
                    &Q_{s,\mu}(q) =  \, (q-1)^{-s-\mu}(1-q^{-\mu})  - q^{-s-\mu}(1 + (q-1)^{-\mu}) - q^{-\mu} + (q-1)^{-\mu}. 
            \label{eq:Mellinint} 
            \end{split}
        \end{equation}
        Integrals~\ref{eq:Mellinint} have to be understood as a principal value integrals as they present divergences in $(0_+, 1_-, 1_+)$. Still, there exists an analyticity strip $\mathcal{I}$ in which the Mellin function is analytical as divergences cancel each others out (see~\cite{zhu2024turbulence}).

        \medbreak\noindent Due to the difference in homogeneity, the 2D ($h_{2D}  =  0$) and 3D ($h_{3D} \neq 0$) cases must be treated separately.
    \subsection{2D case}
        Two-dimensional acoustics is characterised by the peculiar value $h = 0$, leading to a simpler equation:
        \begin{align*}
            \partial_t \hat{A} (s) &= \tau^{-1} \mathcal{W}(s)\hat{A}(s), \\
            \tau &= \frac{\sqrt{6}ac_s}{4\pi V_0^2B},
        \end{align*} 
        that has solutions of the form $\hat{A}(s,t) = f(s) e^{\mathcal{W}(s)t/\tau}$.
        
        \medbreak\noindent The stability of these solutions was previously analysed analytically by Falkovich~\cite{fal1987stability} using a steepest-descent method to compute the inverse Mellin transform of $\hat{A}(s,t)$. In this section, we revisit and refine Falkovich’s calculations and compare the theoretical predictions with numerical simulations.

        \begin{figure}[!htb]
            \centering
            \adjustbox{valign=t}{
                 \begin{minipage}{0.49\textwidth} 
                \includegraphics[width=\textwidth]{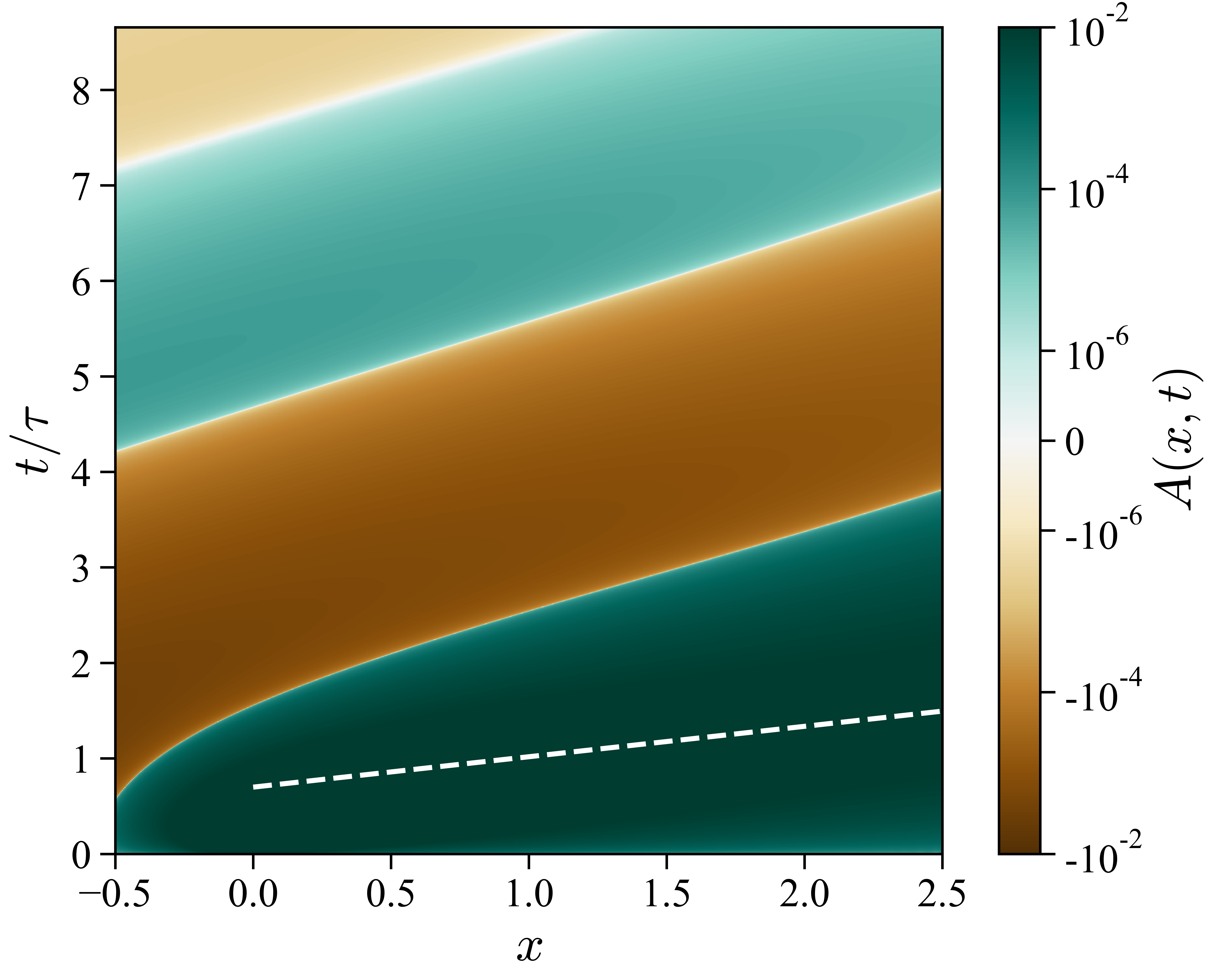}
                \end{minipage}}
            \hfill  
            \adjustbox{valign=t}{\begin{minipage}{0.49\textwidth}
                \includegraphics[width=\textwidth]{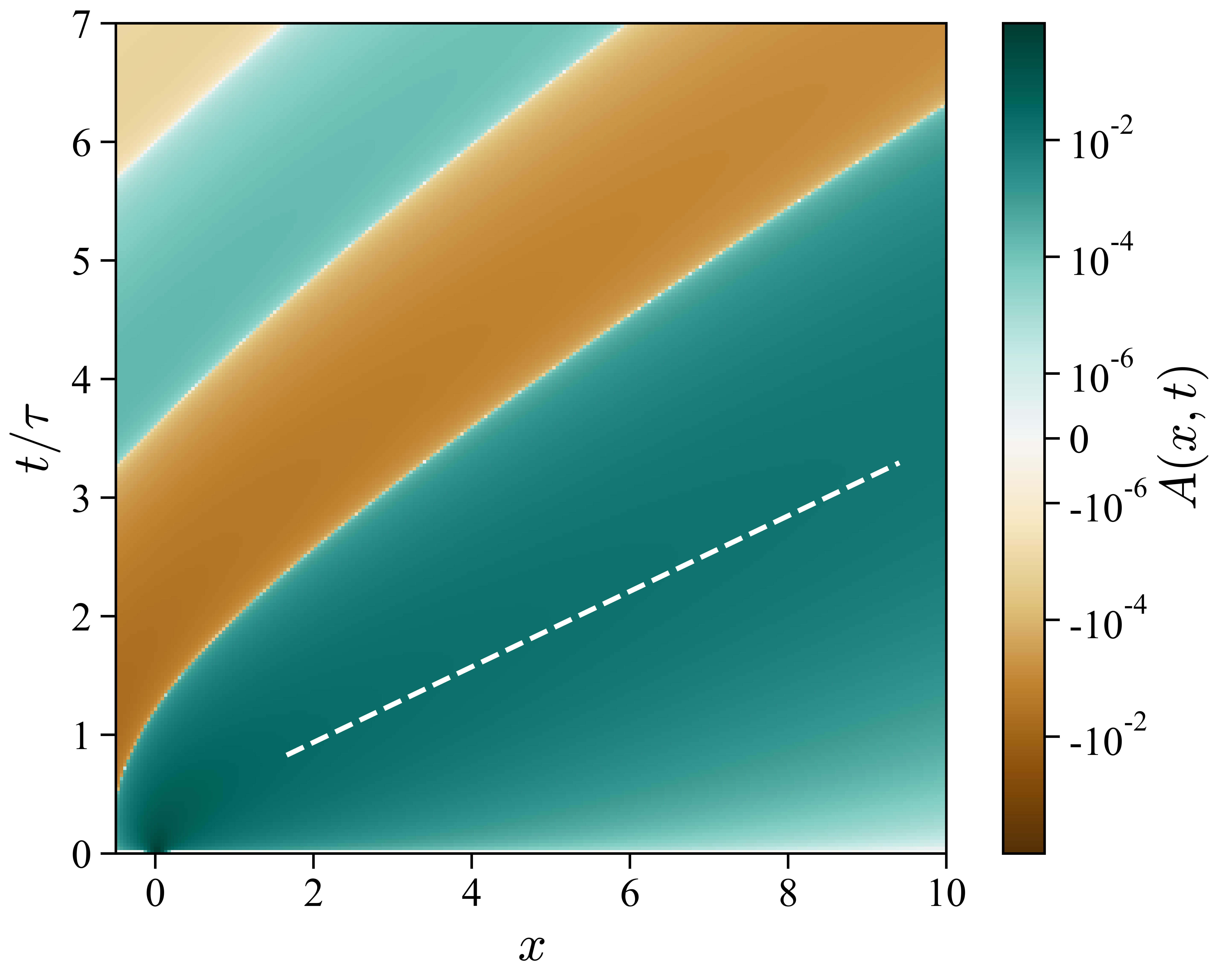}
                \vfill
                \end{minipage}} 
            \put(-509, 10){\resizebox{14pt}{!}{\textbf{(a)}}}
            \put(-251, 10){\resizebox{14pt}{!}{\textbf{(b)}}}
            \put(-444, 10){Numerical Simulation}
            \put(-188, 10){Numerical Integration}
            \caption{Heatmap of the 2D isotropic perturbation $A(x,t)$; $x = \log{k/k_0}$. $\dashed$ $x = V_{0} t$, where $V_{0} = 2\pi$ corresponds to the least damped velocity i.e. $\Gamma'(V_{0}) = 0$. \textbf{(a)} Numerical simulations using WavKinS~\cite{krstulovic2025wavkins}. \textbf{(b)} Numerical calculation of the integral allowing for arbitrarily large $x$, therefore allowing the observation of the combined limits $t \rightarrow +\infty \, ; \, x \rightarrow +\infty$. The numerical integration confirms the observed results of panel \textbf{a} and theoretical predictions of Eq.~\ref{eq:Asympt} (dashed line).}
            \label{fig:2D_heatmap}
        \end{figure}
        \noindent For isotropic perturbations, the Mellin function $\Mellin$ is defined in the strip $-1 < Re(s)< 1$ and can be explicitly computed~\cite{fal1987stability}:
        \begin{equation*}
            \Mellin (s) = 4\pi(1+\frac{s}{2})\tan{\frac{\pi s}{2}},
        \end{equation*}
        \newline The study of the stability of the 2D acoustic KZ solutions then relies on the computation of the inverse Mellin transform of the perturbation $\hat{A}(s,t) = f(s)e^{\mathcal{W}(s)t}$:
        \begin{equation}
            A_k(t) = \int f(s)e^{\mathcal{W}(s)t/\tau}k^{-s}ds. 
            \label{eq:inverseMellin}
        \end{equation}
        In particular, taking initial conditions of the form $\log(f(k)) = -\frac{\log{k} - \log{k_0}}{2\sigma^2}$ for the perturbation~\footnote{while the form of the forcing has no impact on the analytical study, such form is useful for the numerics.}, see inset of Figure.~\ref{fig:KZ}a, one obtains a shape $f(s) \propto k_0^s e^{\frac{\sigma^2}{4}s^2}$. Replacing $f(s)$ in Eq.~\ref{eq:inverseMellin} and introducing $x = \log{k/k_0}$ lead to:
        \begin{equation}
            A_k(t) \propto \int e^\frac{\sigma^2 s^2}{4}e^{\mathcal{W}(s)t/\tau - sx}ds. 
            \label{eq:inverseMellinnum}
        \end{equation}
        Figure~\ref{fig:2D_heatmap}a presents the heatmap of the 2D perturbation, obtained from numerical simulations, highlighting: 
        \begin{enumerate}
            \item propagation in both negative and positive directions,
            \item presence of oscillations in the amplitude.
        \end{enumerate}
        The study of the long-time behaviour of the propagation is thus correlated to the obtention of large inertial range (i.e. obtaining large $|x|$) which represents a real challenge due to the numerical limitations. To complement our observations at larger $|x|$ we performed the numerical integration of the inverse Mellin transform, successfully recovering the heatmap of the perturbation’s propagation for arbitrarily large $|x|$ (Figure~\ref{fig:2D_heatmap}b, spanning more than 10 decades). Both the numerical simulations and integration yield qualitatively similar heatmaps.
\begin{figure}[!htb]
            \centering
            \includegraphics[width=1.\textwidth]{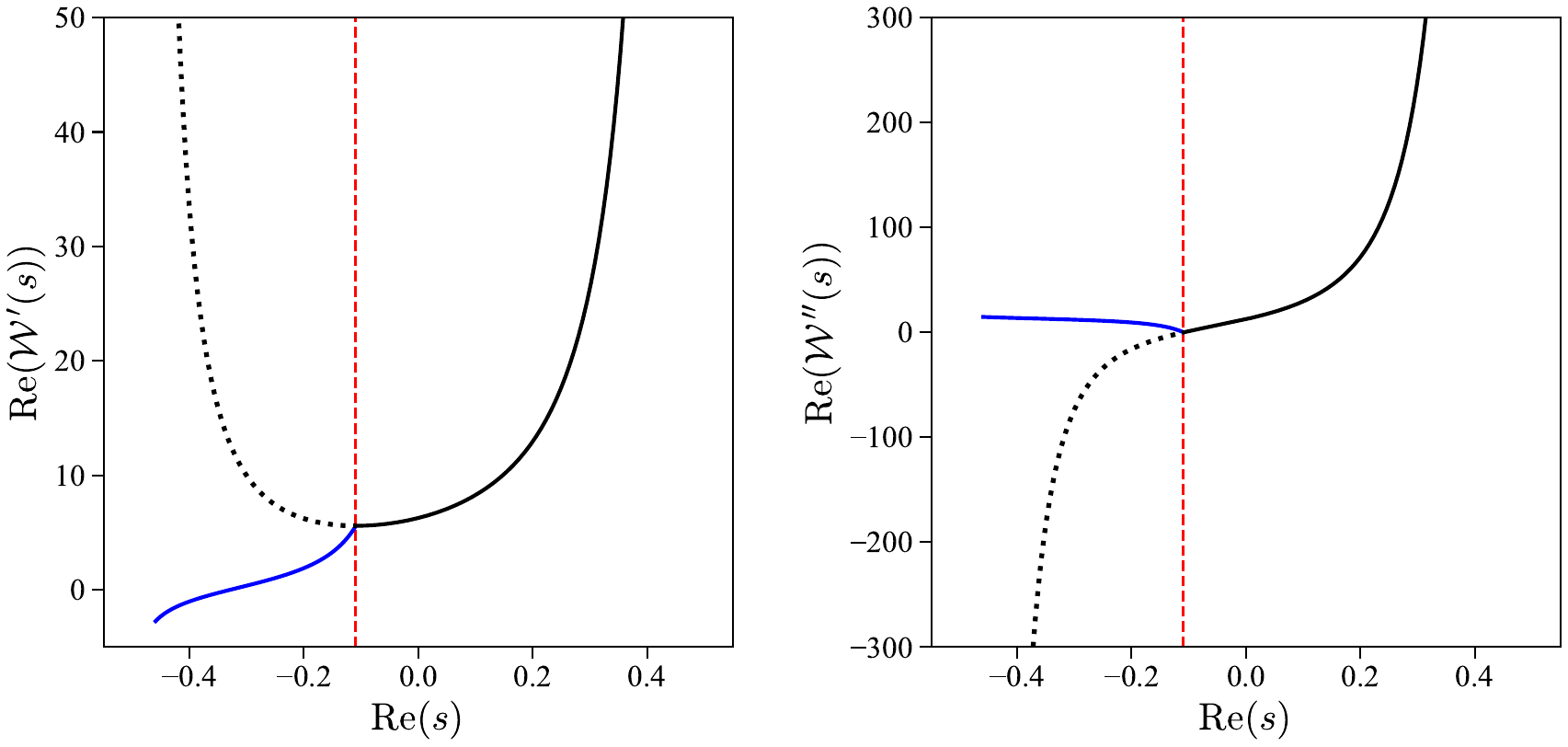}
            \put(-510, 240){\resizebox{14pt}{!}{\textbf{(a)}}}
            \put(-252, 240){\resizebox{14pt}{!}{\textbf{(b)}}}
            \caption{\textbf{(a)} Real part of the derivative of the Mellin function for 2D acoustics. \textbf{(b)} Real part of the second derivative of the Mellin function for 2D acoustics. Both figures share the same legend: $\color{red}{\dashed}$ Critical point $s_0^*\approx-0.11$, defined by $\mathrm{Re}\big(\Mellin''(s_0^*)\big) = 0$, corresponding to the velocity $V_0 = \mathrm{Re}\big(\Mellin'(s_0^*)\big) = 2\pi$. $\dotted$ Negative real saddle points for $V \geq V_0$. $\color{blue}{\full}$ Complex saddle points for $V < V_0$.}
            \label{fig:Mellin}
        \end{figure}
        \medbreak\noindent To explain these observations, we study the long time behavior of $A_k(t)$ that can be estimated using the steepest descent method, relying of the existence of saddle points. We introduce the constant velocity $V \equiv x\tau/t$ such that Eq.~\ref{eq:inverseMellinnum} now reads:
        \begin{equation*}
            A_k(t) =\int h(s)e^{(\mathcal{W}(s)-sV)t/\tau}ds,
        \end{equation*}
        where $h(s) \equiv f(s)/k_0^s$.
        
        \medbreak\noindent The saddle points are defined by the following set of equations:
        \begin{equation}
            \begin{split}
                \mathrm{Re}(\mathcal{W}'(s_V^*)) &= V, \\
                \mathcal{I}m(\mathcal{W}'(s_V^*)) &= 0.
            \label{eq:saddlepoint}
            \end{split}
        \end{equation}
        In the following, we drop the superscript $V$ for simplicity and denote the saddle points as $s^*$.
        \newline\noindent As the function $h(s)e^{(\Mellin(s) - sV)t/\tau}$ is analytical, one can deform the contour of integration to pass through the saddle points allowing for the use of the steepest descent method leading to:
        \begin{align}
            &A_k(t) \approx h(s^*)e^{\Gamma(V) \,t/\tau}\cos{( 2\pi t/T(V))}\int e^{\frac{t}{\tau}\mathrm{Re}\big(\Mellin''(s^*)\big)s^2}ds, \nonumber \\
            &\Gamma(V) = \mathrm{Re}\big(\Mellin(s^*) - s^*V\big), \label{eq:Asympt}\\
            &T(V) =  \frac{2\pi\tau}{\mathcal{I}m\big(\Mellin(s^*) - s^*V\big)}. \nonumber
        \end{align}
        \noindent where $\Gamma(V)$ is the depreciation rate and $T(V)$ the period of oscillations at a given velocity $V$. Note that for real saddle points, ${\mathcal{I}m\big(\Mellin(s^*) - s^*V\big)} = 0$ i.e. $T(V) = +\infty, \; \forall \; V \geq V_0$, thus implying an exponential decay without oscillations. 
        \begin{figure*}[!htb]
            \centering
            \adjustbox{valign=t}{
                \begin{minipage}{0.5\textwidth} 
                    \includegraphics[width=\textwidth]{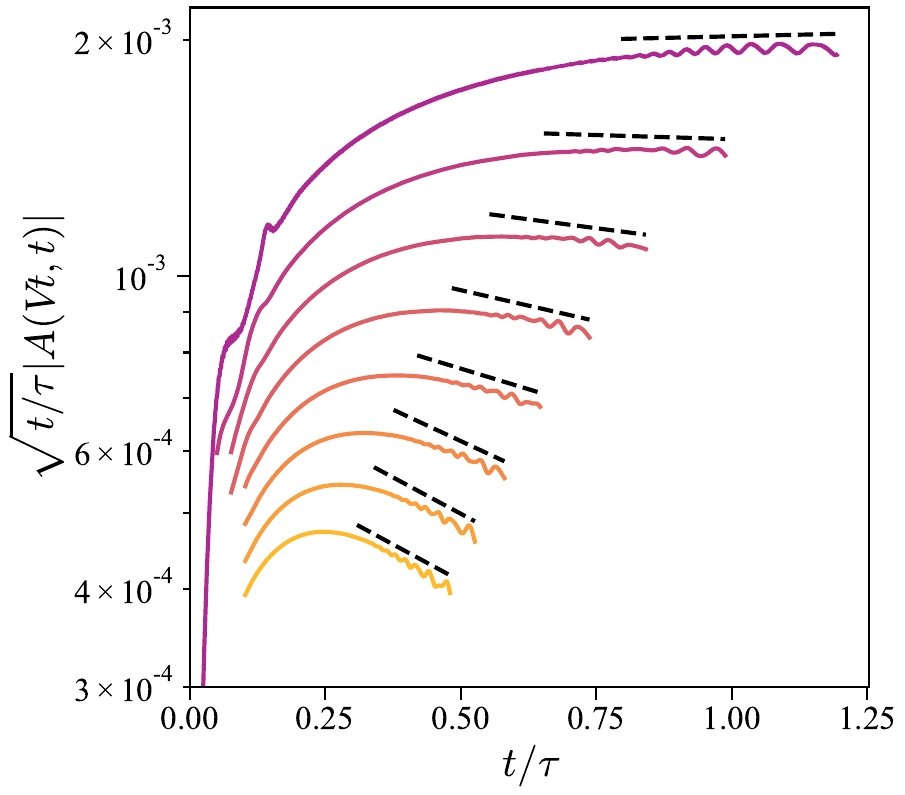}
                \end{minipage}}
                \hfill  
                \adjustbox{valign=t}{\begin{minipage}{0.47\textwidth}
                    \includegraphics[width=\textwidth]{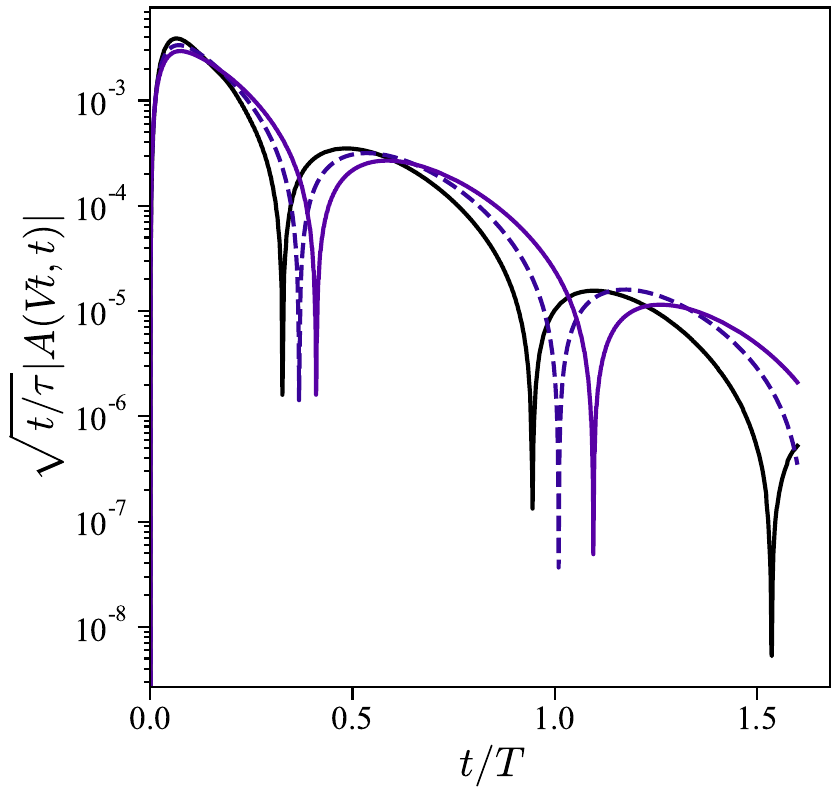}
                    \vfill
                \end{minipage}} 
            \put(-503, 8){\resizebox{14pt}{!}{\textbf{(a)}}}
            \put(-238, 8){\resizebox{14pt}{!}{\textbf{(b)}}}
            \caption{ Slices $|A(Vt, t)|$  for \textbf{(a)} $V \geq V_0$, black dashed lines correspond to the fitting of the damping rate $\Gamma(V)$. \textbf{(b)} $V < V_0$, highlighting oscillations in agreement with the existence of a pair of complex conjugates saddle points. The time is rescaled by the theoretical period $T$ (Eq.~\ref{eq:Asympt}).
            The dashed plot corresponds to the damped trail at fixed $x = \log{(k/k_0)}$. Slices show good agreement with the theoretical predictions of Eq.~\ref{eq:Asympt} and are then used to extract the depreciation rate $\Gamma(V)$.}
            \label{fig:2D_Slice}
        \end{figure*}
        \medbreak\noindent The steepest descent method applied above is, however only valid for saddle points $s^*$ such that the integral $\int_{\sigma -i\infty}^{\sigma +i\infty} e^{\frac{t}{\tau}\mathrm{Re}\big(\Mellin''(s^*)\big)s^2} \, ds$ converges, i.e. when $\mathrm{Re}\big(\Mellin''(s^*)\big) \geq 0$. This condition is not satisfied for real saddle points such that $\mathrm{Re}(s^*) < \mathrm{Re}(s_0^*) \approx 0.11$ (see Figure~\ref{fig:Mellin}b, dotted line), which must therefore be discarded. The asymptotic behavior of the perturbation for all remaining saddle points is given by the following equation:
        \begin{equation}
            A_k(t) \approx \sqrt{\frac{2\pi\tau}{t\,\mathrm{Re}\big(\Mellin''(s^*)\big)}}h(s^*)e^{\Gamma(V) \,t/\tau}\cos{( 2\pi t/T(V))}.
        \label{eq:Asympt_pert}
        \end{equation}
        Solutions of Eq.~\ref{eq:saddlepoint} fall into two distinct categories:  
        \begin{enumerate}
            \item solutions with two real saddle points of opposite sign for  $V \geq \Mellin'(s^*_0) = V_0 = 2\pi$, where $s^*_0 \in \mathbb{R}$ and $\Mellin''(s^*_0) = 0$ (Figure~\ref{fig:Mellin}a, solid and dotted black lines).
            \item solutions with two complex conjugate saddle points for  $V < V_0$ (Figure~\ref{fig:Mellin}a, solid blue line).
        \end{enumerate}

        \medbreak\noindent In particular, a positive velocity allows for the study of the asymptotic behavior $t \to +\infty \; ; \; x \to +\infty$, which depends on the nature of the saddle points. For all $V \geq V_0$, the only accessible saddle point is real and positive, resulting in $T(V) = +\infty$. Figure~\ref{fig:2D_Slice}a presents numerical slices of $|A(Vt, t)|$ for velocities ranging from $V = 0$ (darker) to $V = 14.56$ (lighter), highlighting exponential decay at large times. Note that the small oscillations observed at the largest accessible $t/\tau$ are attributed to numerical errors.
        \begin{figure}[!htb]
            \centering
            \includegraphics[width=0.5\columnwidth]{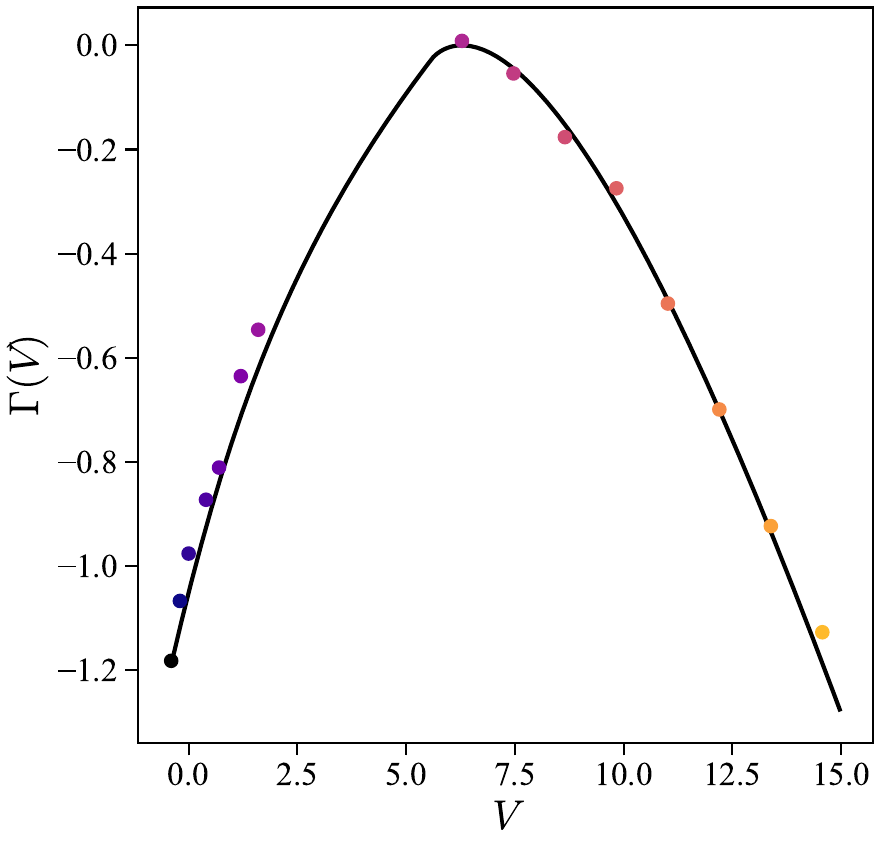}
            \caption{Depreciation rate $\Gamma$ of the asymptotic behavior at large time as a function de the velocity $V$. The black solid line corresponds to the theoretical prediction while colored dots are the numerically extracted (from slices of Figure~\ref{fig:2D_Slice}) values for various velocities showing good agreement with the theoretical predictions of Eq.~\ref{eq:Asympt}.}
            \label{fig:2D_rate}
        \end{figure}
        \medbreak\noindent Any velocity $V < V_0$ is associated with a pair of complex conjugate saddle points, implying the presence of oscillations in the asymptotic behavior. Figure~\ref{fig:2D_Slice}b illustrates the three possible asymptotic regimes:  
        \begin{enumerate}
            \item $0 < V < V_0$ (purple solid line), corresponding to $t \to +\infty \; ; \; x \to +\infty$,
            \item $V = 0$ (dashed line), corresponding to the damped trail left behind at fixed $x$ ,
            \item $V < 0$ (black solid line), corresponding to $t \to +\infty \; ; \; x \to -\infty$.
        \end{enumerate}
        All three cases exhibit oscillations, as highlighted in Figure~\ref{fig:2D_Slice}b, whose periods are in reasonable agreement with the theoretical predictions of Eq.~\ref{eq:Asympt}. Deviations are attributed to limited resolution, which hinders the observation of fully stabilized oscillations.

        \medbreak\noindent In addition, the various slices shown in Figure~\ref{fig:2D_Slice} can be used to extract the depreciation rate $ \Gamma $. Note that extracting the depreciation rate comes with several major numerical hardship. Indeed, for high velocities $V > V_0$, large values of $(x, t)$ must be reached before entering the dissipative range in order to observe the perturbation. In addition, for $V \lesssim V_0$ the period of oscillation $T(V)$ tends to infinity, impeding the proper observation of stabilized oscillation and thus the extraction of the depreciation rate in the range $V \in \big]2.5, 2\pi\big[$
        The resulting depreciation rates are reported in Figure~\ref{fig:2D_rate}, highlighting excellent agreement with the theoretical predictions (solid line) of Eq.~\ref{eq:Asympt}.
        
        \medbreak\noindent We have shown that small isotropic perturbations of the 2D KZ solutions propagate with different velocities $V$. In particular, the propagation is associated with an exponential damping ensuring the stability of the 2D KZ solutions. We now turn to the 3D case, where the presence of non-zero homogeneity $h$ introduces fundamental differences that strongly limit the theoretical analysis.
    
    \subsection{3D case} 
        Three-dimensional acoustics are, unlike the two-dimensional case, characterised by a non-zero homogeneity, $h = -1/2$, in Eq.~\ref{eq:Carleman} so that the Carleman equations reads :
        \begin{equation}
            \partial_tA(s + h) = \Mellin(s) A(s) + \Psi(s+h).
            \label{eq:3D_Carleman}
        \end{equation}
        The study of the stability of the KZ solution is therefore more difficult as the differential equation cannot be solved directly. Nevertheless, Balk \& Zakharov~\cite{balk1998stability} developed a comprehensive theory, relying on the use of the Mellin transform, to analyse the stability of steady-state solutions with respect to small perturbations. Hereafter, we refer to this theory as \textit{Balk-Zakharov (BZ) theory}.
        
        \medbreak\noindent In the following, we analyse the stability of KZ solutions for 3D acoustic waves following the strategy of~\cite{balk1998stability}. As for 2D acoustic waves, the three-dimensional case also appears to be \textit{pathological} due to the properties of the Mellin function $\Mellin$.
        
        \subsubsection{Properties of the Mellin function}
            \begin{figure}[!htb]
                \centering
                \includegraphics[width=1.\textwidth]{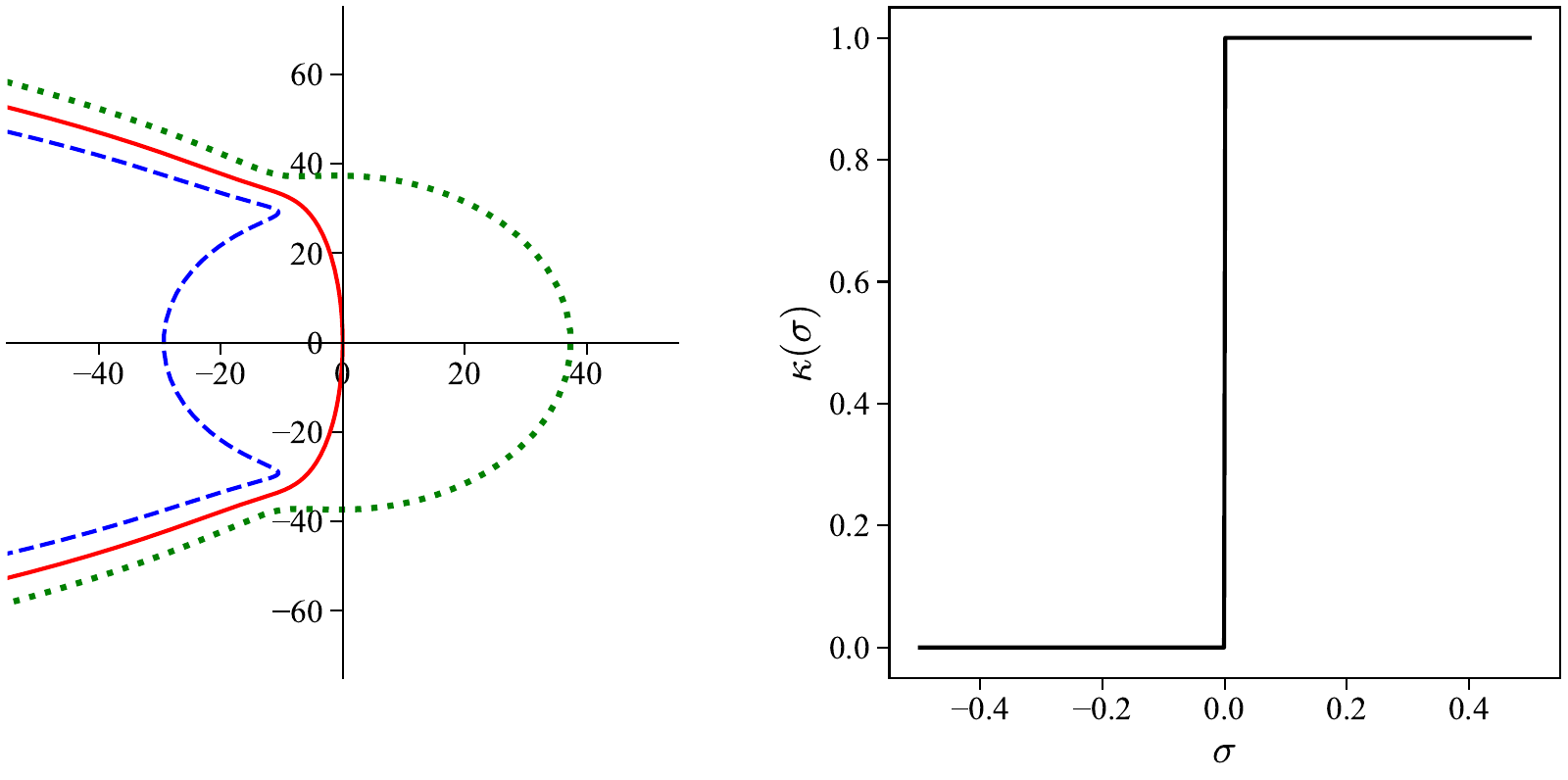}
                \put(-510, 250){\resizebox{14pt}{!}{\textbf{(a)}}}
                \put(-256, 250){\resizebox{14pt}{!}{\textbf{(b)}}}                
                \put(-390, 245){\resizebox{32pt}{!}{$\mathcal{I}m(\Mellin)$}}
                \put(-302, 128){\resizebox{32pt}{!}{$\mathrm{Re}(\Mellin)$}}
                \caption{\textbf{(a)} Path $\mathcal{W}(s)$ in the complex plane, $s = \sigma + i\delta$. $\; \color{blue}{\dashed}$ $\sigma < 0$, $\; \color{red}{\full}$ $\sigma = 0$, $\; \color{darkgreen}{\dotted}$ $\sigma > 0$. \textbf{(b)} Winding number $\kappa(\sigma)$ defined in Eq.~\ref{eq:Winding}.}
                \label{fig:3D_Winding}
            \end{figure}
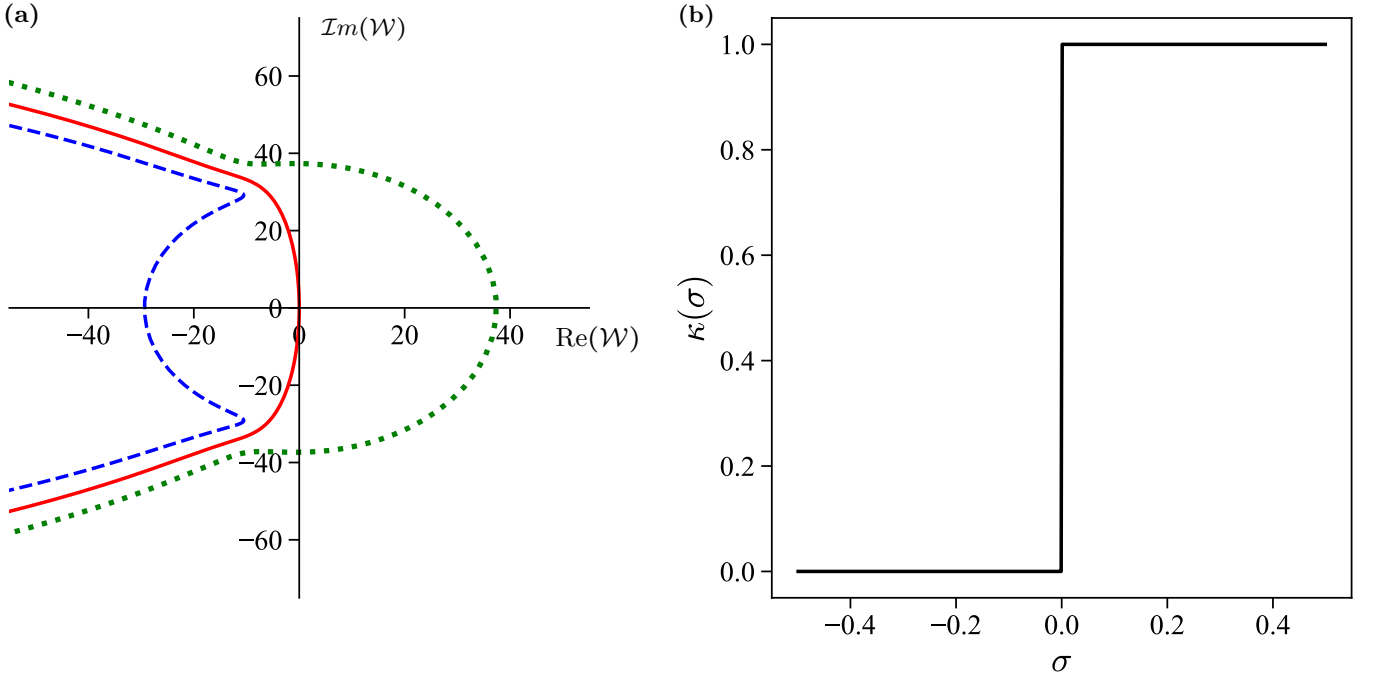
            \noindent The Mellin function for three-dimensional acoustics is defined within the strip $\mathcal{I}_{-1/2,\,1/2}$ and reads:
            \begin{equation*}
                \begin{split}
                    &\forall s\in\mathbb{C} \, / \, -1/2< \mathrm{Re}(s) <1/2, \\
                    &\mathcal{W}(s) = \frac{8}{3}\big( -8 + \frac{24}{(-1+2s)(1+2s)(3+2s)} + \pi^{3/2}\frac{Csc(\pi s) + 2 Csc(2\pi s) + Sec(\pi s)}{\Gamma(-3-s)\Gamma(5/2 + s)}\big).
                \end{split}
            \end{equation*}
            According to BZ theory~\cite{balk1998stability}, all the information about the stability of solutions is encoded inside the Mellin function and its winding number around $0$, $\kappa$, defined as: 
            \begin{equation}
            \label{eq:Winding}
                \begin{split}
                    &\kappa(\sigma) = \frac{1}{2\pi i}\oint_\mathcal{C} \frac{dz}{z}, \\
                    &\mathcal{C} = \big\{ \mathcal{W}(\sigma + i\delta) \;|\; \delta \in \mathbb{R} \big\},
                \end{split}
            \end{equation}
            where the contour $\mathcal{C}$ is the image of the vertical line $\sigma + i\delta$ under the map $\mathcal{W}$, with $\delta \in \mathbb{R}$ (see Figure~\ref{fig:3D_Winding}a).

            \medbreak\noindent The stability is thus determined by (see Appendix~\ref{section:BZ}):
            \begin{itemize}
                \item the existence of a maximal interval $\big[\sigma_-, \sigma_+ \big] \subset \mathbb{R}$ such that:
                \begin{equation*}
                    \forall \sigma \in \big[\sigma_-, \sigma_+ \big], \quad\kappa(\sigma) = 0,
                \end{equation*}
                where $\sigma = \mathrm{Re}(s)$ and $\kappa$ is the winding number (Eq.~\ref{eq:Winding}).
                \item The signs of $\sigma_-$ and $\sigma_+$.
            \end{itemize}
            Note that one always has, for istropic perturbations, $\Mellin(0) = 0$, which can be easily be shown by performing the Zakharov transform to $\Mellin(s)$.
            In addition, the authors would like to add a remark about the necessary condition for stability in the isotropic case mentioned in~\cite{zakharov2012kolmogorov}, namely $\Mellin'(0) > 0$. This condition is clearly satisfied in our case, and in fact for all physically relevant KZ spectra whose prefactor constant is positive (in other words, when the spectral flux has the correct direction). However, this condition is  necessary but not sufficient, which means that the stability result established in the present paper is truly nontrivial.
            
            \medbreak\noindent The existence of such an interval ensures the possibility of constructing a unique solution of the problem through the Wiener-Hopf method. The position of zeroes of the Mellin function with respect to the interval $\big[\sigma_-, \sigma_+ \big]$ determines the stability of the solution~\cite{balk1998stability}.
            
            \medbreak\noindent For 3D acoustics, it can be shown that $\sigma_- = -1/2$ and $\sigma_+ = 0$, as highlighted in Figure~\ref{fig:3D_Winding}. According to BZ theory, we expect the solution to be stable with respect to small isotropic perturbations as $0 \in \big[ \sigma_-, \sigma_+\big]$.
           
        \subsubsection{Building solutions of the Carleman Equation}
            \noindent While directly solving the Carleman equation is not obvious, building solutions of Eq.~\ref{eq:3D_Carleman} remains possible by applying the Laplace transform in time, yielding:
            \begin{equation}
                \lambda A(s-1/2) = \Mellin(s)A(s) + \Psi(s-1/2).
                \label{eq:Carleman_h}
            \end{equation}
            \noindent where $\Psi$ is the Fourier transform of the initial condition.
    
            \medbreak\noindent In particular, a special solution $\mathcal{B}$, called \textit{base function} (solution of Eq.~\ref{eq:Carleman_h} with $\lambda = -1$ and $\Psi \equiv 0$) can be used~\cite{balk1998stability} to build solutions of the general Eq.~\ref{eq:Carleman_h}, through the Wiener-Hopf method. 
            \\ Assuming that the base function $\mathcal{B}$ is known, Eq.~\ref{eq:Carleman_h} is solved as follows:
            \begin{equation*}
            \begin{split}
               & A(s) = \mathcal{B}(s)a(s), \\
                &\mathcal{B}(s) =-\Mellin(s+1/2)\mathcal{B}(s+1/2).
            \end{split}
                \label{eq:AfromBase}
            \end{equation*}
            Substituting in Eq.~\ref{eq:Carleman_h}, one obtains an equation for $a$:
            \begin{equation*}
                \lambda a(s) + a(s+1/2) = \frac{\Psi(s)}{\mathcal{B}(s)}, \, \forall s \in \mathcal{I}_{-1, \, 1/2},
                \label{eq:Eqa}
            \end{equation*}
            that can be solved using a Mellin transform such that $a(s) = \int_{-\infty}^{+\infty}f(x)e^{sx}dx$:
            \begin{equation*}
                \lambda f(x) + f(x)e^{x/2} = P(x).
            \end{equation*}
            So that finally, one has the following set of equations:
            \medbreak\noindent\begin{minipage}{.49\textwidth}
                \begin{align}
                    &A(s) = \mathcal{B}(s)a(s), \\
                    &a(s) = \mathcal{M}[f](s),
                \end{align}
            \end{minipage}%
            \begin{minipage}{.49\textwidth}
                \begin{align}
                    &f(x) = \frac{P(x)}{\lambda+e^{x/2}}, \label{eq:P} \\
                    &P(x) = \mathcal{M}^{-1}[\frac{\Psi}{\mathcal{B}}](x),
                \end{align}
            \end{minipage}
            \newline\noindent where $\mathcal{M}$ (resp. $\mathcal{M}^{-1}$) corresponds to the (resp. inverse) Mellin transform. The Mellin transform of the perturbation $A(s)$ can thus be found by performing several Mellin and inverse Mellin transforms.
            
            \medbreak\noindent It follows that the inverse Mellin transform of the solution can be expressed as:
            \begin{equation}
                A(x,t) = \int_{-\infty}^{+\infty}\mathcal{Z}_2(x-x')\exp{(-te^{x'/2})}P(x')dx',
                \label{eq:Inverse_Pert}
            \end{equation}
            where $\mathcal{Z}_2 = \mathcal{M}^{-1}[\mathcal{B}]$.

            \medbreak\noindent Noticing that $\exp{(-te^{x'/2})} \approx 0$, $\forall (x',t) \in \mathbb{R}^2 \;/ \; |x'| \ll  2 \ln{t}$, one can estimate, from Eq.~\ref{eq:Inverse_Pert}, the large-time behavior of the perturbation using the asymptotic behavior of $P(x)$ as $x\rightarrow -\infty$.
             
            \medbreak\noindent In \textit{BZ theory}, such asymptotical behavior is typically obtained from the \textit{zeros} of the base function $\mathcal{B}$ through the residue theorem (see Appendix~\ref{section:BZ}.d). Yet, the base function for 3D acoustics admits no zeros for $\sigma < \sigma_-$ (see Appendix~\ref{section:BZ}.a), making such an estimate inapplicable. One can thus only rely on the necessity of having a converging convolution integral (see Appendix~\ref{section:BZ}.c) e.g. $P(x) = \mathcal{O}_{-\infty}(e^{x/2})$~\footnote{Understand that $P(x)$ must decrease at least as fast as $e^{x/2}$ in the limit $x\rightarrow -\infty$}.
            \begin{figure*}[!htb]
                \centering
                \adjustbox{valign=t}{
                \begin{minipage}{0.5\textwidth} 
                    \includegraphics[width=\textwidth]{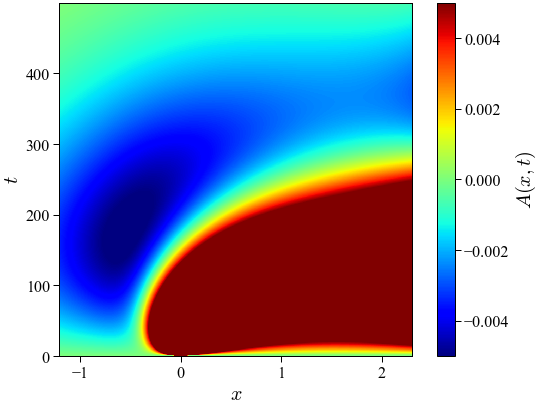}
                \end{minipage}}
                \hfill  
                \adjustbox{valign=t}{
                    \begin{minipage}{0.47\textwidth}
                    \includegraphics[width=\textwidth]{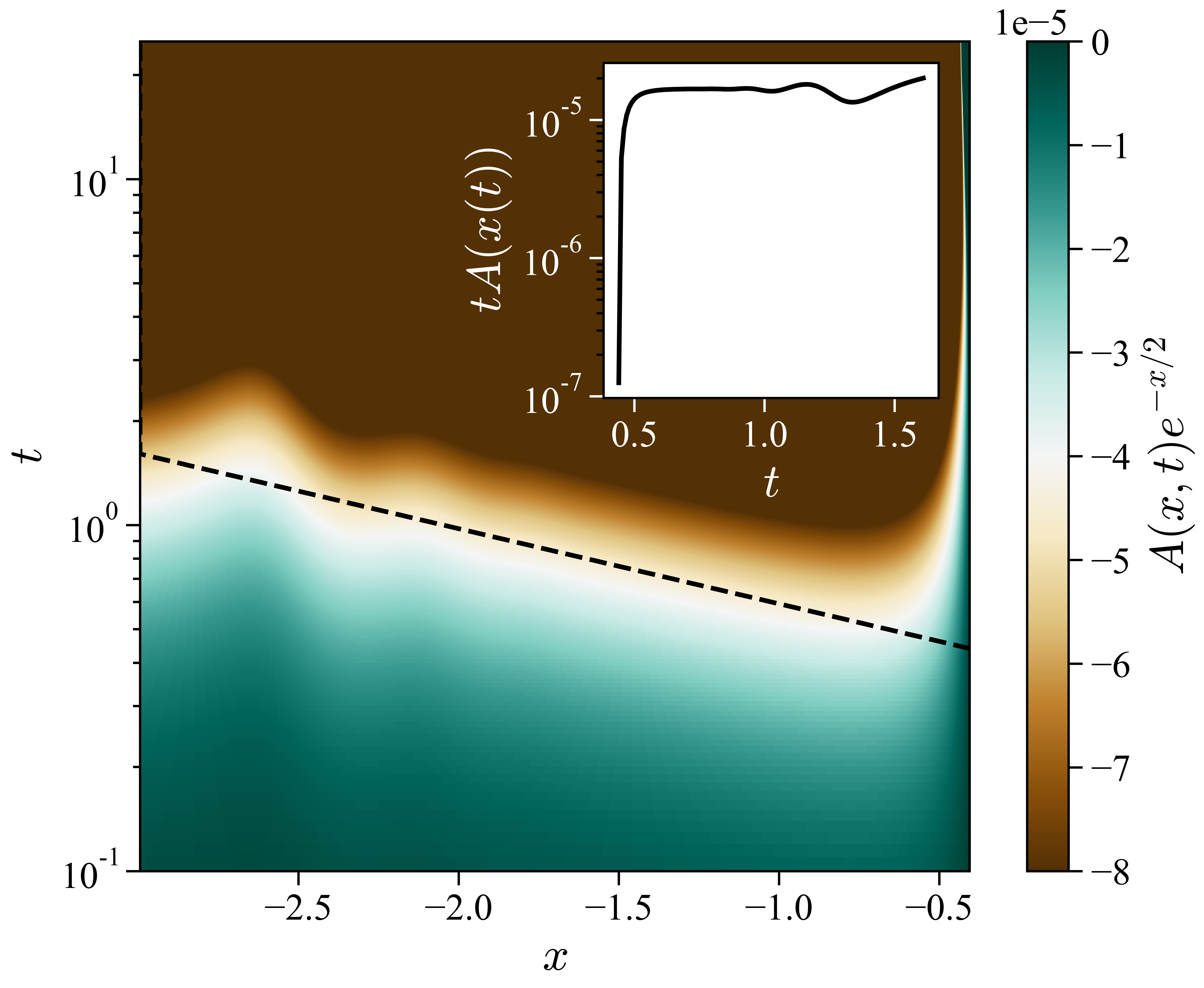}
                    \vfill
                    \end{minipage}} 
                \put(-510, 8){\resizebox{14pt}{!}{\textbf{(a)}}}
                \put(-242, 8){\resizebox{14pt}{!}{\textbf{(b)}}}
                \caption{\textbf{(a)} Heatmap of the 3D perturbation highlighting that a part of the perturbation propagates (i) toward small scales, in agreement with a forward cascade scenario (ii) propagating backwards towards large scales with negative amplitudes. \textbf{(b)} Propagation of the perturbation as a wave running backward. \dashed{} $t = e^{\frac{x_0-x}{2}}$,  where $x_0$ is the first non zero contribution of the initial perturbation $A(x, 0)$. The inset presents the rescaled perturbation $tA(x,t)$ along the dashed line of the main panel confirming the prediction of Eq.~\ref{eq:Selfsim}.}
                \label{fig:3D_heatmap}
            \end{figure*}
            \medbreak\noindent Assuming the slowest possible decay, namely $P(x) = C e^{x/2}$, the perturbation behaves as:
            \begin{equation*}
                A(x,t) \app\limits_{t\rightarrow \infty} C\int_{-\infty}^{+\infty}\mathcal{Z}_2(x-x')[e^{x'/2}\exp{(-te^{x'/2})}]dx',
            \end{equation*}
            \noindent where $C$ is a constant. 
            
            \medbreak\noindent Let $\tau \in \mathbb{R}_+^*$, and consider the perturbation at time $t\tau$, $\forall t > \tau$.
            Performing the change of variable $u = 2\ln(\tau) + x'$ in the above integral yields:
            \begin{equation*}
                A(x,t\tau) \app\limits_{t\rightarrow \infty} C\tau\int_{-\infty}^{+\infty}\mathcal{Z}_2(x + 2\ln{\tau} - u)e^{u/2}\exp{(-te^{u/2})}du.
            \end{equation*}
            The perturbation can thus be expressed as a propagating wave in the negative $x$ direction with an exponential envelope $e^{x/2}$ and possesses the following self-similarity:
            \begin{equation*}
                A(x, t\tau) = A(x+2\ln{\tau}, t)e^{x/2}, 
            \end{equation*}
            \noindent which can be rewritten as:
            \begin{align}
                &\ 
                \left\{\begin{aligned}
                     &A(x,t) = A(x_\tau, \tau)e^{x/2}, \\
                     &x = x_\tau - 2\ln{(t/\tau)},
                \end{aligned}\right.
                 \text \; \forall t \geq \tau > 0
                 \label{eq:Selfsim}
            \end{align}
            \medbreak\noindent In our simulations, the perturbation propagates in two different directions, as highlighted in Figure.~\ref{fig:3D_heatmap}a. Part of it: 
            \begin{enumerate}
                \item propagates backward, towards large scales, with negative amplitude, according to Eq.~\ref{eq:Selfsim} (Figure.~\ref{fig:3D_heatmap}b) further confirmed by the inset, giving the amplitude of the perturbation along the black dashed line of the main figure.
               
                \item propagates toward small scales, in agreement with the forward cascade observed for 3D acoustic waves (Figure~\ref{fig:3D_heatmap}a). While such cascade is not encompassed in the linear theory, \textit{finite capacity} character of 3D acoustics implies that even the smallest amount of energy located at small scales ultimately leads to a direct energy cascade. 
            \end{enumerate}
        
 \section{Conclusion}
    We have analyzed, through a mixture of theory and numerical simulations, the stability of stationary solutions of the WKE in the context of acoustic waves. In particular, we have shown that thermal equilibrium solutions (Rayleigh-Jeans) are stable with respect to small isotropic perturbations. This stability manifests as an exponential decay of the perturbation peak at small times, followed by a broadening that spans the entire spectrum, ultimately leading to a new equilibrium state. The stability of 2D and 3D non-equilibrium (Kolmogorov-Zakharov) solutions has been analyzed using the Mellin function $\Mellin$, which contains all relevant stability information. In particular, a full characterization of 2D perturbation is possible through complex analysis. On the other hand, the stability analysis for 3D KZ solutions is limited by theoretical hardship only allowing us to predict the asymptotic behavior of the perturbation. Nevertheless, our numerical simulations strongly support these theoretical predictions.
    
    \medbreak\noindent While both 2D and 3D KZ solutions exhibit stability under small isotropic perturbations, the perturbation dynamics heavily depends on the dimensionality. Both cases show exponential damping during propagation, but the direction of the propagation is reversed: toward smaller scales in 2D, and toward larger scales in 3D. A forward propagation, not captured by the linear BZ theory, is observed in the 3D case attributed to the the system’s finite capacity property which prevents excessive energy accumulation at large scales.

   \medbreak\noindent Due to the necessity of introducing a small dispersion term to regularize the collision integral, one can not extend this study to next to leading order for anisotropic perturbations, as the regularization breaks the homogeneity of the collision integral. Further work should aim at extending the study to next to leading order anisotropic perturbations and to other physical systems such as the MMT model, or the 3D NLS-WKE inverse cascade.
        
    \section{Acknowledgment}
        This work was funded by the Simons Foundation Collaboration grant Wave Turbulence (Award ID 651471)

\appendix
\section{Locality window}
    \label{section:App_Loca}
    The existence of non-equilibrium steady-state solutions was established in Section~\ref{section:Steady_state} using the Kraichnan-Zakharov transform so as to remap the collision integral onto a unique, dimensionless integral $I(\mu)$ defined as follows:
    \begin{equation*}
        \begin{split}
        &I(\mu) = \int_0^1 \big[q(1 - q)\big]^{D - 1} 
        \left[1 - q^{-y} - (1 - q)^{-y}\right]
        \left[q^{-\mu}(1 - q)^{-\mu} - q^{-\mu} - (1 - q)^{-\mu} \right] \, dq, \\
         &y = 3D - 1 - 2\mu,
        \end{split}
    \end{equation*}
    where we recall that $-\mu$ is the exponent of the isotropic powerlaw solution $n_k = Bk^{-\mu}$.
    \medbreak\noindent However, one must verify the convergence of either $I(\mu)$ or the original collision integral to ensure the applicability of the Kraichnan-Zakharov transform. This interval of convergence is refered to as \textit{locality window}. Let us show that the locality window is indeed given by $\big]2D-2, D+2\big[$~\cite{griffin2022energy, zhu2024turbulence}.

    \medbreak\noindent First, note that $\forall \mu$, $I(\mu)$ is invariant under the change of variable $q \rightarrow 1-q$, such that one only needs to check the convergence at the lower-bound of the integral. We therefore examine the asymptotic behavior of each term in the integrand near $q \to 0$, trivially one has:
    \begin{equation*}
            \left[q(1-q)\right]^{D-1}  \underset{q \to 0}{\sim} q^{D-1}.
    \end{equation*}
    To treat the term $\left[q^{-\mu}(1 - q)^{-\mu} - q^{-\mu} - (1 - q)^{-\mu} \right]$, 
    one must expand $(1 - q)^{-\mu}$ to the next-to-leading order. Using the Taylor expansion around $q = 0$, we write $(1 - q)^{-\mu} =1+\mu q + \mathcal{O}_{0}(q)$, leading to the following equivalent:
    \begin{equation*}
        \left[q^{-\mu}(1 - q)^{-\mu} - q^{-\mu} - (1 - q)^{-\mu} \right] \underset{q \to 0}{\sim} q^{1-\mu}
    \end{equation*}
    Finally, one must be careful when deriving the asymptotic behavior of $\left[1 - q^{-y} - (1 - q)^{-y}\right]$ as it depends on the sign of $y$:
    \begin{equation*}
         \left[1 - q^{-y} - (1 - q)^{-y}\right] \underset{q \to 0}{\sim} \begin{cases}
             q^{-y}  &y > 0, \\
             q   &y < 0.
         \end{cases}
    \end{equation*}
    The integrand then behaves as:
    \begin{equation*}
        \big[q(1 - q)\big]^{D - 1} \left[1 - q^{-y} - (1 - q)^{-y}\right]\left[q^{-\mu}(1 - q)^{-\mu} - q^{-\mu} - (1 - q)^{-\mu} \right] \underset{q \to 0}{\sim} q^{-\beta}
    \end{equation*}
    with $\beta$ given by:
        \begin{equation*}
         \beta = \begin{cases}
         -\mu -1 +2D  &y > 0, \\
             \mu -1 - D  &y \leq 0.
         \end{cases}
    \end{equation*}
    where we have substituted $y = 3D - 1 - 2\mu$ into the previously obtained equivalent. Gathering the previous inequalities, the integral thus converges if and only if $\beta < 1$ leading to the condition:
    \begin{equation*}
        2D-2 <\mu < D+2 
    \end{equation*}
    effectively recovering the announced locality window.

\section{Rayleigh-Jeans damping rate}
    \label{section:Ansatz_RJ}
    We consider the following evolution equation for the perturbation $A(k,t)$, written in frequency variables:
    \begin{equation*}
        \begin{split}
            \partial_t A(\omega_k,t) &= \frac{1}{n_k^0} \int \left( \mathcal{A}^{\bm{k}}_{\bm{1},\bm{2}} - 2\mathcal{A}^{\bm{1}}_{\bm{k},\bm{2}} \right) \, d\omega_1 \, d\omega_2, \\
            \mathcal{A}^{\bm{k}}_{\bm{1},\bm{2}} &= \frac{(k_1 k_2)^{D-1}}{|\partial_{k_1} \omega_1 \, \partial_{k_2} \omega_2|} |W^{\bm{k}}_{\bm{1},\bm{2}}|^2 \delta(\omega^k_{1,2}) \left[ A_1 n_1^0(n_2^0 - n_k^0) + A_2 n_2^0(n_1^0 - n_k^0) - A_k n_k^0(n_1^0 + n_2^0) \right], \\
            |W^{\bm{k}}_{\bm{1},\bm{2}}|^2 &= 2\pi |V^{\bm{k}}_{\bm{1},\bm{2}}|^2 \langle \delta^{\bm{k}}_{\bm{1},\bm{2}} \rangle, \\
            \delta(\omega^k_{1,2}) &= \delta(\omega_k - \omega_1 - \omega_2),
        \end{split}
    \end{equation*}
    where $A_i \equiv A(\omega_i,t)$, $n_k^0$ is a stationary solution of the WKE, and $\langle \delta^{\bm{k}}_{\bm{1},\bm{2}} \rangle$ denotes the integration over angular variables of the wavevector resonance condition.
    
    \medbreak\noindent In the following, we suppose the perturbation $A(\omega_k,t)$ to be smooth and sharply peaked around $\omega_p$ such that:
    \begin{equation}
        A(\omega_k,t) = 
        \begin{cases}
            A(\omega_k,t), &\quad |\omega_k - \omega_p| < \epsilon, \\
            0,           &\quad \text{otherwise},
        \end{cases}
        \label{eq:Perturb}
    \end{equation}
    where $\epsilon \ll \omega_p$ defines the width of the localized perturbation. In the following, we thus restrict the analysis to the case $\omega_k = \omega_p$.

    \medbreak\noindent The evolution equation can then be rewritten as:
    \begin{align*}
        \partial_t A(\omega_p,t) =& -A(\omega_p,t) \! \int \!\frac{(k_1 k_2)^{D-1}}{|\partial_{k_1} \omega_1 \, \partial_{k_2} \omega_2|} \!\left[ |W^p_{1,2}|^2 \delta(\omega^p_{1,2})(n_1^0 + n_2^0) + 2|W^1_{p,2}|^2 \delta(\omega^1_{p,2})(n_2^0 - n_1^0) \right] d\omega_1 \, d\omega_2 \\
        &+ \frac{1}{n_p^0} \int \frac{(k_1 k_2)^{D-1}}{|\partial_{k_1} \omega_1 \, \partial_{k_2} \omega_2|} \left( \Lambda_{1,2}^{p} - 2\Lambda_{p,2}^{1} \right) \, d\omega_1 \, d\omega_2,
    \end{align*}
    where $\Lambda^p_{1,2}$ is defined as:
    \begin{equation*}
        \Lambda^p_{1,2} = |W^p_{1,2}|^2 \delta(\omega^p_{1,2}) \left[ A_1 n_1^0(n_2^0 - n_p^0) + A_2 n_2^0(n_1^0 - n_p^0) \right].
    \end{equation*}
    We now analyze the behavior of the integral:
    \begin{equation*}
        J= \frac{1}{n_p^0}\int \frac{(k_1k_2)^{D-1}}{|\partial_{k_1}\omega_1\partial_{k_2}\omega_2|}\big(\Lambda_{1,2}^{p}-2\Lambda_{p,2}^{1}\big)\,d\omega_1\,d\omega_2.
    \end{equation*} 
    To do so, and to fix the idea, we consider only the term:
    \begin{equation*}
        I =\frac{1}{n_p^0}\int \frac{(k_1k_2)^{D-1}}{|\partial_{k_1}\omega_1\partial_{k_2}\omega_2|}|W^p_{1,2}|^2\delta(\omega^p_{1,2})A_1n_1^0(n_2^0 - n_p^0)\,d\omega_1\,d\omega_2.
    \end{equation*}
    
    \medbreak\noindent Integrating over the resonance constraint leads to:
    \begin{equation*}
        I = \sum_{\omega_2^*} \int \frac{(k_1 k_2^*)^{D-1}}{|\partial_{k_1} \omega_1 \, \partial_{k_2} \omega_2^*|} |W^p_{1,2^*}|^2 A(\omega_1) n_1^0(n_{2^*}^0 - n_p^0) \, d\omega_1,
    \end{equation*}
    where the sum runs over the solutions $\omega_2^*(\omega_1)$ of the frequency resonance condition. For clarity, we drop the sum in the remainder.

    \medbreak\noindent Inserting the definition Eq.~\ref{eq:Perturb} and keeping only leading-order contributions, we obtain:
    \begin{equation*}
        I(\omega_p) \approx A(\omega_p) \left[ \frac{(k_1 k_2^*)^{D-1}}{|\partial_{k_1} \omega_1 \, \partial_{k_2} \omega_2^*|} |W^p_{1,2^*}|^2 n_1^0(n_{2^*}^0 - n_p^0) \right]_{\omega_1 = \omega_p}.
    \end{equation*}
    The frequency resonance condition implies:
    \begin{equation*}
    \omega_2^* = \omega_p - \omega_1 \xrightarrow[\omega_1 \to \,\omega_p]{} 0.
    \end{equation*}
    Therefore, we conclude that:
    \begin{equation*}
    I(\omega_p) \approx 0,
    \end{equation*}
    under the reasonable assumption that the kernel
    \begin{equation*}
    \frac{(k_1 k_2^*)^{D-1}}{|\partial_{k_1} \omega_1 \, \partial_{k_2} \omega_2^*|} |W^p_{1,2^*}|^2
    \end{equation*}
    remains finite in the infrared limit $\omega_2^* \to 0$.
    
    \medbreak\noindent The same argument applies to all terms in $J$, which are thus negligible. Hence, at leading order, the evolution equation reduces to:
    \begin{equation*}
        \partial_t A(\omega_p,t) \approx -\alpha(\omega_p) A(\omega_p,t),
    \end{equation*}
    where the damping rate $\alpha(\omega_p)$ is given by:
    \begin{equation*}
        \alpha(\omega_p) = \int \frac{(k_1 k_2)^{D-1}}{|\partial_{k_1} \omega_1 \, \partial_{k_2} \omega_2|} \left[ |W^p_{1,2}|^2 \delta(\omega^p_{1,2})(n_1^0 + n_2^0) + 2|W^1_{p,2}|^2 \delta(\omega^1_{p,2})(n_2^0 - n_1^0) \right] \, d\omega_1 \, d\omega_2.
    \end{equation*}
    
    \medbreak\noindent In particular, for acoustic waves, one has $\omega_k = c_s k$ and $W^k_{1,2} = \sqrt{c_sC_D}$ (with $C_D$ defined in Eq.~\ref{eq:Cd}) such that:
    \begin{equation*}
        \alpha(k_p) = c_sC_D \int (k_1 k_2)^{D-1} \left[ \delta(\omega^p_{1,2})(n_1^0 + n_2^0) + 2 \delta(\omega^1_{p,2})(n_2^0 - n_1^0) \right] \, dk_1 \, dk_2.
    \end{equation*}
    
    \medbreak\noindent Finally, for the Rayleigh-Jeans distribution:
    \begin{equation*}
        n_k^0 = \frac{E^{RJ}}{\mathcal{S}_D \omega_k},
    \end{equation*}
    where $\mathcal{S}_D$ is the volume of a $D$-dimensional hypersphere of radius $k_{\max}$ and $E^{RJ}$ is the total energy of the RJ state, the damping rate reads:
    \begin{equation*}
        \alpha(k_p) = \frac{E^{RJ} C_D}{c_s \mathcal{S}_D} k_p \left( 2 \int_{k_p}^{k_{\max}} \left[ k_1(k_1 - k_p) \right]^{D-2} \, dk_1 + \int_{0}^{k_p} \left[ k_1(k_p - k_1) \right]^{D-2} \, dk_1 \right).
    \end{equation*}
    
\section{A quick tour of the BZ theory}
    As deriving a comprehensive explanation of the BZ theory would be out of the scope of this appendix. We only propose an overview of the fundamental steps necessary to the understanding of the paper and refer the reader to Chapter $2-3$ by Balk \& Zakharov~\cite{balk1998stability}.
    \label{section:BZ}
    \subsection{Carleman type equation and zero winding interval}
        \noindent Performing a Mellin transform in $k$ and a Laplace transform in $t$ to the linearized WKE leads to the following Carleman-type equation:
        \begin{equation}
            \lambda A(s+h) = \Mellin(s)A(s) + \Psi(s+h),
            \label{eq:Carleman_th_app}
        \end{equation}
        where $A(s)$ is the Mellin transform of the small perturbation $A_k \equiv \delta n_k / n_k^0$, $\Mellin$ is the Mellin function, analytic in a strip $\mathcal{I}_{ab}$ (i.e. $\mathrm{Re}(s) \in ]a,b[$) and $h$ is the homogeneity of the linearized collision integral. As for 3D acoustics, one has $(a,b,h) = (-1/2, 1/2,-1/2$), in the following, we consider $h < 0$, discarding the trivial $h=0$ case associated with 2D acoustics. The positive $h$ case can be obtained by changing $h \to -h \; ; x \to -x$.
        
        \medbreak\noindent Solutions of Eq.~\ref{eq:Carleman_th_app} can be obtained using the Wiener-Hopf method that will not be discussed here. In particular, if there exists a strip $\strip$ in which the winding number $\kappa(\mathrm{Re}(s)) =0\; \forall s\in \strip$ then Eq.~\ref{eq:Carleman_th_app} admits a unique solution $A_0$ in the strip $\strip$. Note that for any linearized kinetic equation, the winding number is a monotonically non-decreasing function implying that $\mathcal{W}$ has neither zeros nor poles in $\strip$. In the following, we suppose that such strip exists.

        \medbreak\noindent If $a<\sigma_-$, the solution $A_0$ can be extended to the left of the zero-winding number strip by iterating Eq.~\ref{eq:Carleman_th_app}. \noindent Similarly, by translating the Carleman equation~\ref{eq:Carleman_th_app}, one obtains:
        \begin{equation*}
            A(s) = \frac{\lambda A(s+h) - \Psi(s+h)}{\Mellin(s)}.
        \end{equation*}
        This equation can then be iterated to extend the solution to the right of the strip $\strip$. The extended solution is well defined in the strip $\mathcal{I}_{a+h, b}$.
     
    \subsection{Homegeneous Carleman equation and base function}
        \noindent The homogeneous Carleman equation is obtained from Eq.~\ref{eq:Carleman_th_app} by setting $\Psi \equiv 0$ so that it reads:
        \begin{equation}
            \lambda A(s+h) = \Mellin(s)A(s).
            \label{eq:Carleman_h_app}
        \end{equation}
        As in the previous section, any solution to the homogeneous equation in $\strip$ can be extended to the strip $\mathcal{I}_{a+h, b}$. In particular, such solution possesses no poles in $\mathcal{I}_{a+h, \sigma_+}$ and its zeros are determined by those of the Mellin function, such that:
        \medbreak\noindent \textit{If $q \in \mathcal{I}_{a, \sigma_-}$ is a zero of the Mellin function $\Mellin(s)$. Then, the extended solution has zeros at the points \begin{center}
        $q + h,\, q+2h,\, q+3h,\, \dots$
        \end{center}
        }
        \noindent Similarly, by translating the Homegeneous Carleman equation, one obtains:
        \begin{equation*}
            A(s) = \frac{\lambda A(s+h)}{\Mellin(s)}.
        \end{equation*}
        This equation can then be iterated to extend the solution to the right of the strip $\strip$. As a result, the extended solution $A$ is meromorphic in the strip $\mathcal{I}_{a+h, b}$, since zeros of $\Mellin$ located to the right of $\strip$ correspond to poles of $A$:
        \medbreak\noindent \textit{$p \in \mathcal{I}_{\sigma_+, b}$ is a zero of the Mellin function $\Mellin(s)$. Then, the extended solution has poles at the points \begin{center}
        $p,\, p-h,\, q-2h,\, \dots$
        \end{center}
        }
        \noindent The position of the zeros and poles of the extended solution $A$ can be summarized by the following graph:
        \bigbreak
        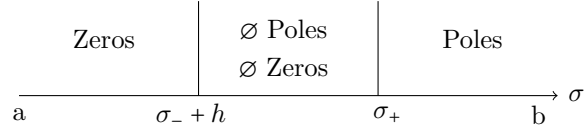
\begin{figure}[!htb]
            \centering
            \begin{tikzpicture}[scale=0.50]
                \draw [->] (6.25,11.25) -- (20.5,11.25);
                \draw (11,11.25) -- (11,13.75);
                \draw (15.75,11.25) -- (15.75,13.75);
                \node [font=\normalsize] at (21,11.25) {$\sigma$};
                \node [font=\normalsize] at (10.8,10.75) {$\sigma_- + h$};
                \node [font=\normalsize] at (16,10.75) {$\sigma_+$};
                \node [font=\normalsize] at (6.25,10.75) {a};
                \node [font=\normalsize] at (20,10.75) {b};
                \node [font=\normalsize] at (8.5,12.75) {Zeros};
                \node [font=\normalsize] at (18.25,12.75) {Poles};
                \node [font=\normalsize] at (13.25,12) {$\emptyset$ Zeros};
                \node [font=\normalsize] at (13.25,13) {$\emptyset$ Poles};
            \end{tikzpicture}
            \caption{Position, $\sigma = \mathrm{Re}(s)$, of zeros and poles of the extended solution $A$ of Eq.~\ref{eq:Carleman_h_app}.}
            \label{fig:PolesandZeros}
        \end{figure}
        
        \medbreak\noindent In particular, a special solution $\mathcal{B}$, called \textit{base function}, solution of Eq.~\ref{eq:Carleman_h_app} with $\lambda = -1$, can be used~\cite{balk1998stability} to build solutions of the general Eq.~\ref{eq:Carleman_th_app}, through the Wiener-Hopf method. 
        \medbreak\noindent Assuming that the base function $\mathcal{B}$ is known, Eq.~\ref{eq:Carleman_th_app} is solved as follows:
        \begin{equation*}
            A(s) = \mathcal{B}(s)a(s).
            \label{eq:AfromBase_app}
        \end{equation*}
        Substituting in Eq.~\ref{eq:Carleman_th_app}, one obtains an equation for $a$:
        \begin{equation*}
            \lambda a(s) + a(s-h) = \frac{\Psi(s)}{\mathcal{B}(s)}, \, \forall s \in \mathcal{I}_{a+h, b},
            \label{eq:Eqa_app}
        \end{equation*}
        that can be solved using a Mellin transform such that $a(s) = \int_{-\infty}^{+\infty}f(x)e^{sx}dx$:
        \begin{equation*}
            \lambda f(x) + f(x)e^{-hx} = P(x).
        \end{equation*}
        So that finally, one obtains the following set of equations $\forall s \, \in \strip$:
        \medbreak\noindent\begin{minipage}{.49\textwidth}
                \begin{align}
                    &A(s) = \mathcal{B}(s)a(s), \\
                    &a(s) = \mathcal{M}[f](s),
                \end{align}
            \end{minipage}%
            \begin{minipage}{.49\textwidth}
                \begin{align}
                    &f(x) = \frac{P(x)}{\lambda+e^{x/2}}, \label{eq:P_app} \\
                    &P(x) = \mathcal{M}^{-1}[\frac{\Psi}{\mathcal{B}}](x),
                \end{align}
            \end{minipage}
        \newline\noindent where $\mathcal{M}$ (resp. $\mathcal{M}^{-1}$) corresponds to the (resp. inverse) Mellin transform. The solution is then extended to the whole strip $\mathcal{I}_{a+h, b}$ through iteration of the Carleman equation (see Section~\ref{section:BZ}.a).
    
    \subsection{Existence of the Mellin transform}
        \noindent In order to reach Eq.~\ref{eq:Carleman_th_app} it is necessary for the following convolution to exist at all time:
        \begin{equation}
            (\mathcal{U}*A)(x) = \int_{-\infty}^{+\infty} \mathcal{U}(x-x')A(x',t)dx',
            \label{eq:convolution_app}
        \end{equation}
        where the kernel $\mathcal{U}$ possesses the following properties:
        \begin{equation*}
         \mathcal{U}(x) =
         \left\{\begin{matrix} 
         \mathcal{O}_{-\infty}(e^{-ax}),
         \\
         \mathcal{O}_{+\infty}(e^{-bx}).
        \end{matrix}\right.
        \end{equation*}
        
        \noindent Convolution~\ref{eq:convolution_app} exists if the perturbation satisfies the conditions:
        \begin{equation*}
            \forall t, \,
            A(x,t) =
            \begin{cases}
            \mathcal{O}_{-\infty}(e^{-\sigma_2x}), \, \sigma_2 <b, \\
            \mathcal{O}_{+\infty}(e^{-\sigma_1x}), \, \sigma_1 >a.
            \end{cases}
        \end{equation*}
        $A_0$ being solution of Eq.~\ref{eq:Carleman_th_app} in $\strip$ implies that the above conditions are fulfilled in this region, with $\sigma_1 = \sigma_-$ and $\sigma_2 = \sigma_+$. Therefore, the extension to the left of the strip leads to $\forall t, \ A(x,t) = \mathcal{O}_{-\infty}(e^{-(a+h)x})$ so that we have:
        \begin{equation*}
            \forall t, \,
            A(x,t) =
            \begin{cases}
            \mathcal{O}_{-\infty}\!\left(e^{-(a+h)x}\right), \\
            \mathcal{O}_{+\infty}\!\left(e^{-\sigma_1 x}\right), \qquad \sigma_1 > a .
            \end{cases}
        \end{equation*}
        
    \subsection{Behavior of the perturbation}
        \subsubsection{Asymptotic behavior at \texorpdfstring{$\bm{|x| \to +\infty}$}{eq1}}
         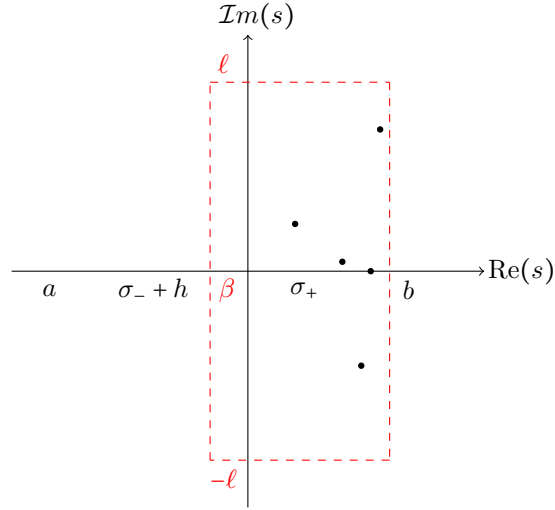
\begin{figure}[!htb]
                \centering
                \begin{tikzpicture}[scale=0.50]
                    \tikzstyle{every node}=[font=\LARGE]
                    \draw [->] (5,9) -- (17.5,9);
                    \draw [->] (11.25,2.75) -- (11.25,15.25);
                    \node [font=\normalsize] at (18.5,9) {$\mathrm{Re}(s)$};
                    \node [font=\normalsize] at (11.5,15.75) {$\mathcal{I}m(s)$};
                    \draw [ color={rgb,255:red,255; green,0; blue,0}, dashed] (10.25,4) -- (10.25,14);
                    \draw [ color={rgb,255:red,255; green,0; blue,0}, dashed] (10.25,14) -- (15,14);
                    \draw [ color={rgb,255:red,255; green,0; blue,0}, dashed] (15,14) -- (15,4);
                    \draw [ color={rgb,255:red,255; green,0; blue,0}, dashed] (15,4) -- (10.25,4);
                
                    \node [font=\normalsize, color={rgb,255:red,255; green,0; blue,0}] at (10.6,14.5) {$\ell$};
                    \node [font=\normalsize,  color={rgb,255:red,255; green,0; blue,0}] at (10.6, 3.5) {$-\ell$};
                    
                    \node [font=\normalsize, color={rgb,255:red,255; green,0; blue,0}] at (10.7,8.5) {$\beta$};
                    \node [font=\normalsize] at (15.5,8.5) {$b$};
                    \node [font=\normalsize] at (6,8.5) {$a$};
                    \node [font=\normalsize] at (8.75,8.5) {$\sigma_- + h$};
                    \node [font=\normalsize] at (12.75,8.5) {$\sigma_+$};
                
                    \filldraw[black] (13.75, 9.25,0) circle (2pt) node[]{};
                    \filldraw[black] (12.5, 10.25,0) circle (2pt) node[]{};
                    \filldraw[black] (14.25, 6.5,0) circle (2pt) node[]{};
                    \filldraw[black] (14.5, 9,0) circle (2pt) node[]{};
                    \filldraw[black] (14.75, 12.75,0) circle (2pt) node[]{};
                \end{tikzpicture}
                \caption{Finite size rectangular contour to perform the residue theorem. The black dots correspond to poles of $A$ that only exist for $\mathrm{Re}(s) \geq \sigma_+$ (see Figure~\ref{fig:PolesandZeros}).}
                \label{fig:Contour}
            \end{figure}
            \noindent The asymptotic behavior of $A(x,t)$ for $x \to +\infty$ can obtained by performing the inverse Mellin transform :
            
            \begin{equation*}
                A_\lambda(x) = \int_{\alpha -i\infty}^{\alpha +i\infty}A_\lambda(s)e^{-sx}ds,
            \end{equation*}
            Integrating along the rectangular contour $\mathcal{C}$ of Figure~\ref{fig:Contour} and using the residue theorem one obtains, taking $\ell\to +\infty$, 
            \begin{equation*}
                A(x,t) \app\limits_{x\to +\infty} \sum\limits_{q} \text{res}(A(q, t))e^{-qx},
            \end{equation*}
            where the sum runs over the poles of the perturbation $A(s,t)$. Note that the integrals along the horizontal segments vanish as $\ell \to \infty$, while the contribution from the vertical segment at $s = \beta < 0$ vanishes in the limit $x\to+\infty$.
           
            \noindent In particular, the sum is dominated by the poles with smallest real part such that
            \begin{equation*}
                A(x,t) \app \sum\limits_{\mathrm{Re}(q) = \sigma_+} \text{res}(A(q, t))e^{-qx}, \, x \to +\infty.
            \end{equation*}
            As a result, the assumptotic behavior of $A$ is given by:
            \begin{equation}
                \forall t, \quad A(x, t) =
                \begin{cases}
                    \mathcal{O}_{-\infty}(e^{-(a+h)x}), & \text{as } x \to -\infty, \\
                    \mathcal{O}_{+\infty}(e^{-\sigma_+x}), & \text{as } x \to +\infty,
                \end{cases}
                \label{eq:Bounds_app}
            \end{equation}
            
        \subsubsection{Time evolution of the perturbation}
            \noindent It can be shown~\cite{balk1998stability} that solutions of Eq.~\ref{eq:Carleman_th_app} are given by:
            \begin{equation*}
                 A_\lambda(x) = \mathcal{Z}^{-1}(\lambda + e^{-hx})^{-1}\mathcal{Z}[\Phi],
            \end{equation*}
            where $\mathcal{Z}$ and $\mathcal{Z}^{-1}$ are two operators defined from the base function $\mathcal{B}(s)$, solution of the homogeneous Carleman equation (see Section~\ref{section:BZ}.b), such that:
            \begin{equation}
                \forall (f,x), \quad
                \begin{cases}
                    \mathcal{Z}[f](x) = (\mathcal{Z}_1 * f)(x), \\
                    \mathcal{Z}^{-1}[f](x) = (\mathcal{Z}_2 * f)(x),
                \end{cases}
                \label{eq:Operators_app}
            \end{equation}
            \noindent where $\mathcal{Z}_1 = \mathcal{M}^{-1}[\mathcal{B}^{-1}]$ and $\mathcal{Z}_2 = \mathcal{M}^{-1}[\mathcal{B}]$.

            \medbreak\noindent As both operators are independent of time, one can invert the Laplace transform to obtain:
            \begin{equation*}
                \begin{split}
                    &A(x,t) =  \mathcal{Z}^{-1}[ \theta_t(x)P(x)], \\
                    &\theta_t(x) = \exp{(-te^{-hx})},
                \end{split}
            \end{equation*}
            corresponding to a wave traveling in the negative $x$ direction. 
            
            \medbreak\noindent Combining the above equations and Eq.~\ref{eq:P_app} \& \ref{eq:Operators_app}, one obtains the following expression for the perturbation:
            \begin{equation*}
                A(x,t) = \int_{-\infty}^{+\infty}\mathcal{Z}_2(x-x')\theta_t(x')P(x')dx'
            \end{equation*}
            In addition, $\forall (x,t) \in \mathbb{R}^2\  / x\ll  x_t = h^{-1}\ln{t}, \, \theta_t(x) \approx0$ meaning that for large time one can write
            \begin{equation*}
                \theta_t(x) \app \limits_{t \to +\infty} \exp{(-te^{-hx})}(1-H(x-x_t)),
            \end{equation*} 
            where $H$ denotes the Heaviside step function. 
            \newline\noindent Therefore, one can replace $P$ (Eq.~\ref{eq:P_app}) by its asymptotic behavior at $x \to -\infty$ that can be estimated via the residue theorem using a contour similar to that in Figure~\ref{fig:Contour}:
            \begin{equation*}
                P(x) \approx
                \left\{\begin{matrix} 
                \sum\limits_{q}C(q)e^{-qx},& \, x \to -\infty
                \\
                \mathcal{O}_{+\infty}(e^{-bx}), &
                \end{matrix}\right. 
            \end{equation*}
            where the sum runs over the zeros of $\mathcal{B}$. The large time behavior of the perturbation can then be estimated by:
            \begin{equation*}
                \begin{split}
                    A(x,t) &\app\limits_{t\to +\infty} \sum\limits_{q}C(q)\int_{-\infty}^{x_t}\mathcal{Z}_2(x-x')[e^{-qx'}\exp{(-te^{-hx'})}]dx',
                            \\
                    &\app\limits_{t\to +\infty}\sum\limits_{q} A_q(x,t).
                \end{split}
            \end{equation*}
            The perturbation can thus be interpreted as a superposition of waves $A_q$, each propagating in the negative $x$-direction with an envelope that depends on the corresponding pole $q$. In particular, each wave $A_q$ exhibits the following self-similarity property:
            \begin{equation*}
                A_q(x, t\tau) = A_q(x-h^{-1}\ln{\tau}, t)\tau^{-q/h}, 
            \end{equation*}
            \noindent which can be rewritten as:
            \begin{align}
                &\ 
                \left\{\begin{aligned}
                    &A_q(x,t) = A_q(x_\tau, \tau) e^{-qx}, \\
                    &x = x_\tau + h^{-1}\ln{(t/\tau)},
                    \end{aligned}\right.
                \text \; \forall t \geq \tau
                \label{eq:Selfsim_app}
            \end{align}
            In particular, the dominant envelope of the perturbation $A$ is determined by the zeros of $\mathcal{B}$ with the smallest real part, corresponding to an envelope of the form $e^{-(\sigma_- + h)x}$ (see Figure~\ref{fig:PolesandZeros}).

        \subsubsection{Stability of solution}
            \noindent The stability of the solution can be analyzed using Eqs.~\ref{eq:Bounds_app} and~\ref{eq:Selfsim_app}. From Eq.~\ref{eq:Bounds_app}, we observe that for the perturbation to remain bounded as $x \to +\infty$, it is necessary that $\sigma_+ \geq 0$. On the other hand, Eq.~\ref{eq:Selfsim_app} implies that the perturbation decays as it propagates if $\sigma_- + h < 0$. 
            \medbreak \noindent
            Taken together, these two conditions ensure the stability of the KZ solution, which holds if:
            \begin{equation}
                \sigma_- + h < 0 \leq \sigma_+,
            \end{equation}
            a criterion that can be summarized as $\kappa(0) = 0$.

\bibliography{apssamp} 

\end{document}